\documentclass[sigconf,10pt]{acmart}

\makeatletter
\newif\if@restonecol
\makeatother

\usepackage[lined,boxed,vlined,ruled]{algorithm2e}
\usepackage{balance}

\usepackage{hyperref}

\usepackage{multicol}

\usepackage{listings}
\usepackage{framed}
\usepackage{xcolor}
\colorlet{shadecolor}{gray!20}

\newcommand{\eat}[1]{}
\usepackage{latexsym}
\usepackage{amsfonts}
\usepackage{amsmath}
\usepackage{amssymb}
\usepackage{color}
\usepackage{colortbl}
\usepackage{epsfig}
\usepackage{xspace}
\usepackage{graphicx}
\usepackage{subfigure}

\usepackage{paralist}
\usepackage{enumerate}
\usepackage{pifont}
\usepackage{ifthen}
\newboolean{isTechReport}
\setboolean{isTechReport}{true}   
\newcolumntype{L}[1]{>{\raggedright\let\newline\\\arraybackslash\hspace{0pt}}m{#1}}
\newcolumntype{C}[1]{>{\centering\let\newline\\\arraybackslash\hspace{0pt}}m{#1}}
\newcolumntype{R}[1]{>{\raggedleft\let\newline\\\arraybackslash\hspace{0pt}}m{#1}}

\newcommand{\hi}[1]{\vspace{.25em} \noindent {\bf #1}\xspace}

\newcommand{\qtonly}{\ensuremath{\texttt{TrainingOnly}}\xspace}

\newcommand{\qconly}{\ensuremath{\texttt{ClusterOnly}}\xspace}
\newcommand{\qronly}{\ensuremath{\texttt{TransOnly}}\xspace}

\newcommand{\random}{\ensuremath{\texttt{Interleave-Random}}\xspace}
\newcommand{\local}{\ensuremath{\texttt{Interleave-Greedy}}\xspace}
\newcommand{\greedy}{\ensuremath{\texttt{Interleave-Greedy}}\xspace}
\newcommand{\globalone}{\ensuremath{\texttt{Interleave-Global-1}}\xspace}
\newcommand{\globalk}{\ensuremath{\texttt{Interleave-Global-k}}\xspace}
\newcommand{\globalkcorr}{\ensuremath{\texttt{Interleave-Global-k-Corr-b}}\xspace}
\newcommand{\emec}{\ensuremath{\texttt{EMEC}}\xspace}

\newcommand{\globalten}{\ensuremath{\texttt{Interleave-Global-10}}\xspace}

\newcommand{\globalkcorrfive}{\ensuremath{\texttt{Interleave-Global-10-Corr-5}}\xspace}

\newcommand{\pub}{\ensuremath{\texttt{Pub}}\xspace}
\newcommand{\addr}{\ensuremath{\texttt{Address}}\xspace}
\newcommand{\product}{\ensuremath{\texttt{Product}}\xspace}

\newcommand{\budget}{\ensuremath{B}\xspace}

\newcommand{\budgetq}{\ensuremath{\mathcal{B}}\xspace}
\newcommand{\costq}{\ensuremath{\mathcal{C}}\xspace}
\newcommand{\benefitq}{\ensuremath{\mathcal{B}}\xspace}
\newcommand{\pro}{\ensuremath{\mathcal{P}}\xspace}

\newcommand{\nuq}{\ensuremath{\mathcal{N}}\xspace}

\newcommand{\qex}{\ensuremath{q}\xspace}

\newcommand{\qt}{\ensuremath{q_T}\xspace}

\newcommand{\qc}{\ensuremath{q_C}\xspace}
\newcommand{\qr}{\ensuremath{q_R}\xspace}

\newcommand{\Qt}{\ensuremath{\mathbb{Q}_T}\xspace}

\newcommand{\Qc}{\ensuremath{\mathbb{Q}_C}\xspace}
\newcommand{\Qr}{\ensuremath{\mathbb{Q}_R}\xspace}
\newcommand{\Qs}{\ensuremath{\mathbb{Q}^*}\xspace}
\newcommand{\Qall}{\ensuremath{\mathbb{Q}}\xspace}
\newcommand{\Qb}{\ensuremath{\mathbb{Q}^b}\xspace}

\newcommand{\dataD}{\ensuremath{\mathbb{D}}\xspace}

\newcommand{\cov}{\ensuremath{{\texttt{cov}}}\xspace}
\newcommand{\pre}{\ensuremath{{\texttt{accuracy}}}\xspace}
\newcommand{\uti}{\ensuremath{{\texttt{utility}}}\xspace}

\usepackage{epsfig}
\usepackage{multirow}
\usepackage{url}

\newcommand{\mourad}[1]{\footnote{\textcolor{cyan}{Mourad: #1}}}

\definecolor{orange}{HTML}{FF7F00}

\newcommand{\ra}{\rightarrow}

\newcommand{\bi}{\begin{itemize}}
\newcommand{\ei}{\end{itemize}}

\newcommand{\be}{\begin{enumerate}}
\newcommand{\ee}{\end{enumerate}}
\newcommand{\beqn}{\begin{eqnarray*}}
\newcommand{\eeqn}{\end{eqnarray*}}

\newcommand{\stitle}[1]{\vspace{1ex}\noindent{\bf #1}}

\newcommand{\ie}{{i.e.,}\xspace}
\newcommand{\eg}{{e.g.,}\xspace}
\newcommand{\wrt}{\emph{w.r.t.}\xspace}
\newcommand{\aka}{{a.k.a.}\xspace}

\newcounter{ccc}

\newcommand{\eop}{\hspace*{\fill}\mbox{$\Box$}\vspace{1ex}}     
\newcounter{example}
\renewcommand{\theexample}{\arabic{example}}
\newenvironment{example}{
        \vspace{1ex}
        \refstepcounter{example}
        {\noindent\bf Example \theexample:}}{
        \eop}

\newcommand{\nthesection}{\arabic{section}}
\newcounter{theorem}
\renewcommand{\thetheorem}{\arabic{theorem}}
\newcounter{prop}[section]

\newcounter{lemma}[section]
\renewcommand{\thelemma}{\nthesection.\arabic{theorem}}
\newcounter{cor}
\renewcommand{\thecor}{\arabic{theorem}}

\newcounter{alg}[section]
\renewcommand{\thealg}{\nthesection.\arabic{alg}}

\newcounter{arule}
\renewcommand{\thearule}{\arabic{arule}}

\newcounter{claim}
\renewcommand{\theclaim}{\arabic{claim}}

\DeclareMathOperator*{\argmax}{arg\,max}

\makeatletter
    \newcommand\figcaption{\def\@captype{figure}\caption}
    \newcommand\tabcaption{\def\@captype{table}\caption}
\makeatother

\definecolor{shadecolor1}{RGB}{230,230,230}

\definecolor{shadecolor1}{RGB}{255, 114, 118}

\usepackage[all]{nowidow}

\begin{document}

\pagestyle{plain}
\pagenumbering{roman}

\newpage

\setcounter{page}{1}
\pagenumbering{arabic}

\title{Technical Report: Optimizing Human Involvement for Entity Matching and Consolidation}

\author{
Ji~Sun\ding{93}~~~~Dong~Deng\ding{168}~~~~Ihab~Ilyas\ding{169}~~~~Guoliang Li\ding{93} \\
Samuel~Madden\ding{170}~~~~Mourad~Ouzzani\ding{171}~~~~Michael~Stonebraker\ding{170}~~~~Nan~Tang\ding{171}
}
\affiliation{
\ding{93}Tsinghua~University~~~~\ding{168}Rutgers~University~~~~\ding{169}University~of~Waterloo~~~~\ding{170} MIT~CSAIL~~~\ding{171}QCRI
}
\email{
sun-j16@mails.tsinghua.edu.cn; liguoliang@tsinghua.edu.cn; ilyas@uwaterloo.ca
}
\email{
{dongdeng,madden,stonebraker}@csail.mit.edu;{mouzzani,ntang}@hbku.edu.qa
}

\begin{abstract}
An end-to-end data integration system requires human feedback in several phases, including collecting training data for entity matching, debugging the resulting clusters,  confirming transformations applied on these clusters for data standardization,  and finally, reducing each cluster to a single, canonical representation (or ``golden record'').  The traditional wisdom is to sequentially apply the human feedback, obtained by asking specific questions, within some budget in each phase.  However, these questions are highly correlated; the answer to one can influence the outcome of any of the phases of the pipeline. Hence, interleaving them has the potential to offer significant benefits.

In this paper, we propose a human-in-the-loop framework that interleaves different types of questions to optimize human involvement. We propose benefit models to measure the quality improvement from asking a question, and cost models to measure the human time it takes to answer a question. We develop a question scheduling framework that judiciously selects questions to maximize the accuracy of the final golden records.  Experimental results on three real-world datasets show that our holistic method significantly improves the quality of golden records from 70\% to 90\%, compared with the state-of-the-art approaches.
\end{abstract}



\maketitle


\vspace{-.5em}
\section{Introduction}
\label{sec:introduction}

An end-to-end data integration system typically involves the following phases: obtaining training data to construct an entity matching (EM) module, executing this module to find duplicate records, and constructing clusters by grouping duplicate records;  debugging the clusters; transforming the variant values into the same format; reducing each cluster into to a canonical record (\aka golden record) by entity consolidation (EC), which is the final output.


Based on experience with more than 100 real-world data integration projects at a well established data integration company, Tamr\footnote{https://www.tamr.com}, 
we note several common trends:


%

\hi{(a)} 	Human involvement is needed throughout the integration process, in three distinct tasks: 
\be
\item {\em Training Rule Validation.} Different from the candidate pairs obtained through blocking, 
the training pairs for the EM classifier should have more accurate labels. However, it is usually infeasible to generate training data one pair of records at a time. Instead, training data can be constructed from a collection of human-written or machine-generated rules~\cite{DBLP:journals/pvldb/WangLYF11,DBLP:journals/pvldb/MollZPMSG17}, such as ``{\it if the Jaccard similarity of the addresses of two companies is larger than 0.8, then the two companies represent the same entity}''.  
To ensure that these rules generate high-quality training data, it is necessary to validate them using human input along with a sample of the training data they generate. 

\item {\em Cluster Validation.} Running an EM model to find duplicates is usually followed by a clustering algorithm that groups all duplicates into clusters.
Some clusters must be validated by a human to ensure their correctness. 

\item {\em Transformation Validation.} Each cluster must be reduced to a single golden record. 
Within a cluster, the same entity may be represented using different values.
One way to consolidate them is to 
transform them into the same format using transformation rules, which are generated from current clusters, such as ``\textsf{CS} $\rightarrow$ \textsf{Computer Science}''. These rules  also need to be validated by humans. 
\ee
\vspace{-.5em}
\hi{(b)} These three tasks are typically executed sequentially.
However, as we show below, these tasks are highly correlated  and  
interleaving them can offer significant benefits. 

\hi{(c)} At scale, human involvement must be optimized, since human time dominates the cost of data integration projects.  There is no hope of exhaustively checking all these tasks.

To improve the entire process, we study the problem of {\em optimizing human involvement in entity matching and consolidation}. Our goal is to optimize the human involvement  by interleaving the aforementioned three tasks.


\subsection{Opportunities} 

\begin{figure}[t!]\vspace{-2em}
\begin{center}
\centering
\tabcaption{A Raw Table $\mathbb{D}$
\label{tbl:table}}
\hspace*{-1em}
\includegraphics[width=\columnwidth]{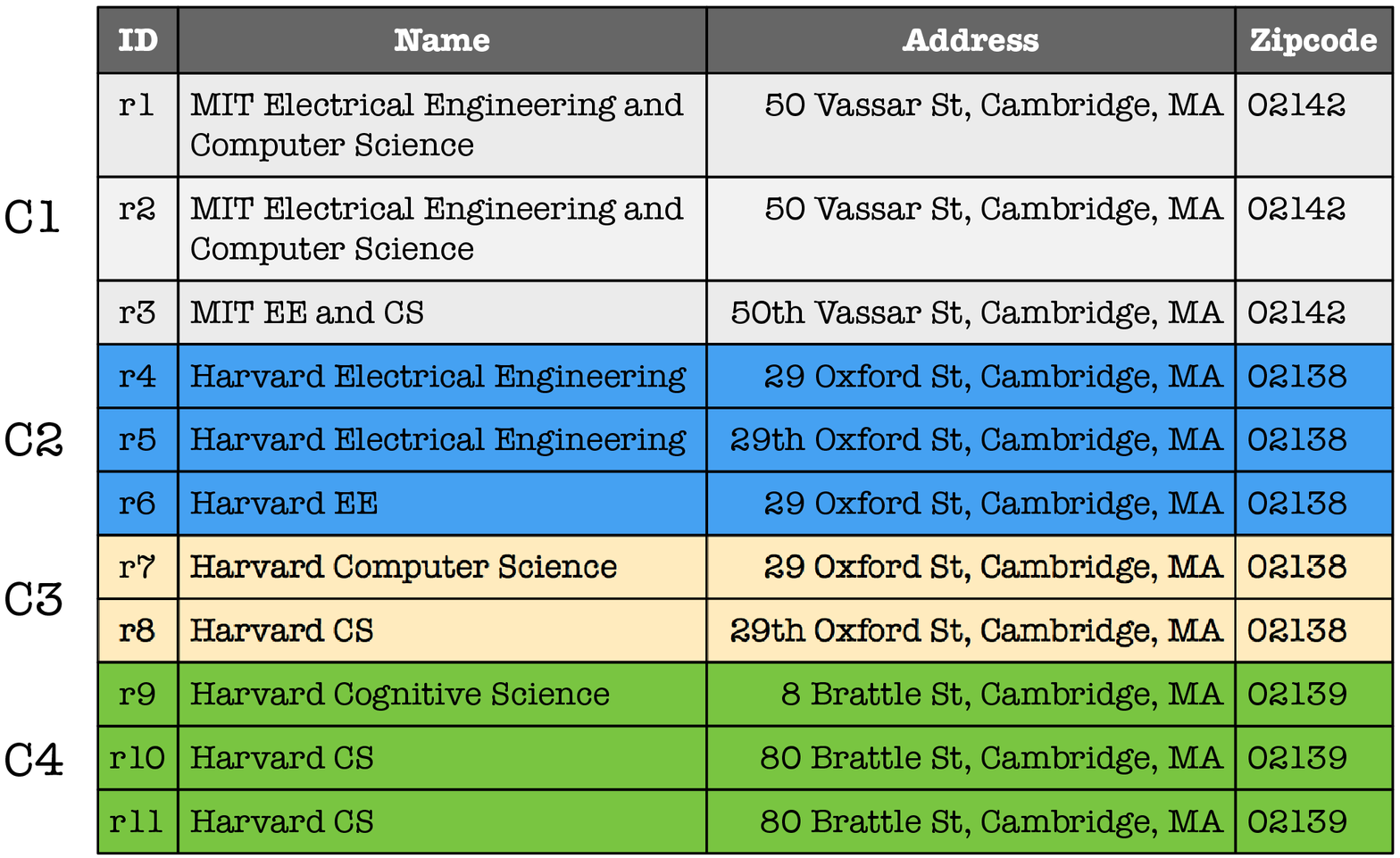}
\vspace{-1.5em}
\end{center}
\end{figure}

\begin{figure}[t!]
\begin{center}
\centering
\tabcaption{Golden Records of Table $\mathbb{D}$\label{gr_example}
\label{tbl:grecords}}
\hspace*{-1em}\includegraphics[width=\columnwidth]{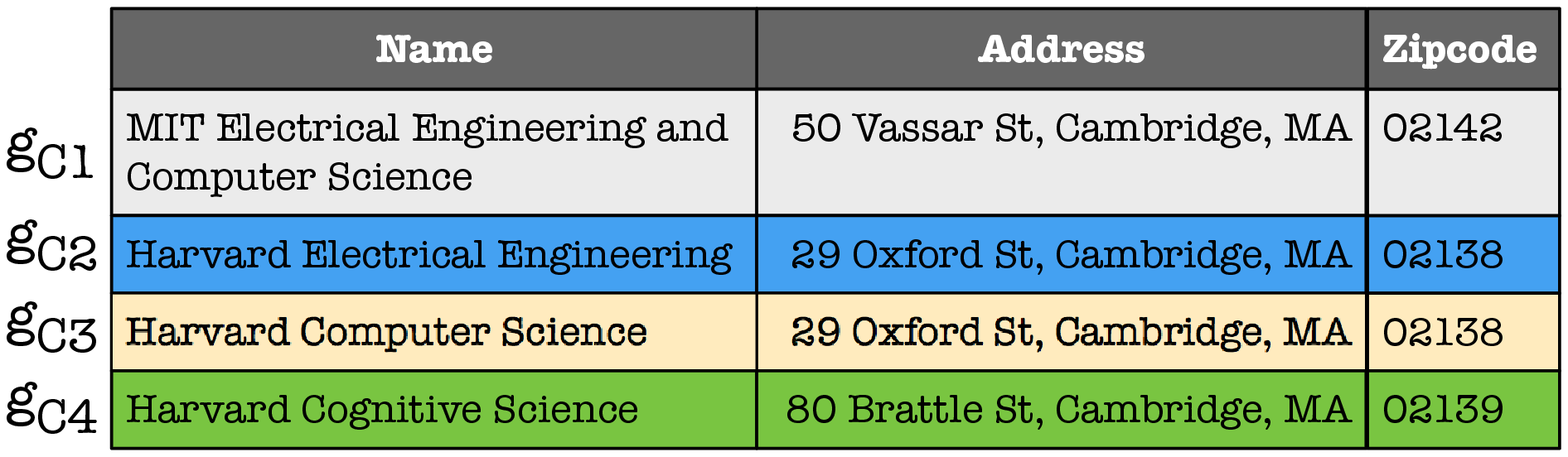}
\vspace{-1.25em}
\end{center}
\end{figure}

Let us first show through an example what happens if we run the aforementioned phases sequentially. Consider Table~\ref{tbl:table} with 11 records that refer to 4 real-world entities (\ie clusters $\{C_1, C_2, C_3, C_4\}$). Their ground truths are shown in Table~\ref{tbl:grecords}, where $g_C{_i}$ represents the ground truth for $C_i$ ($i\in [1, 4]$).



\begin{figure}[!t]\vspace{-2em}
 \begin{center}
\subfigure[{\bf EM $\rightarrow$ DT$\rightarrow$ EC}]{
     \label{fig:ex:sequential1}
     \includegraphics[width=1.05\columnwidth]{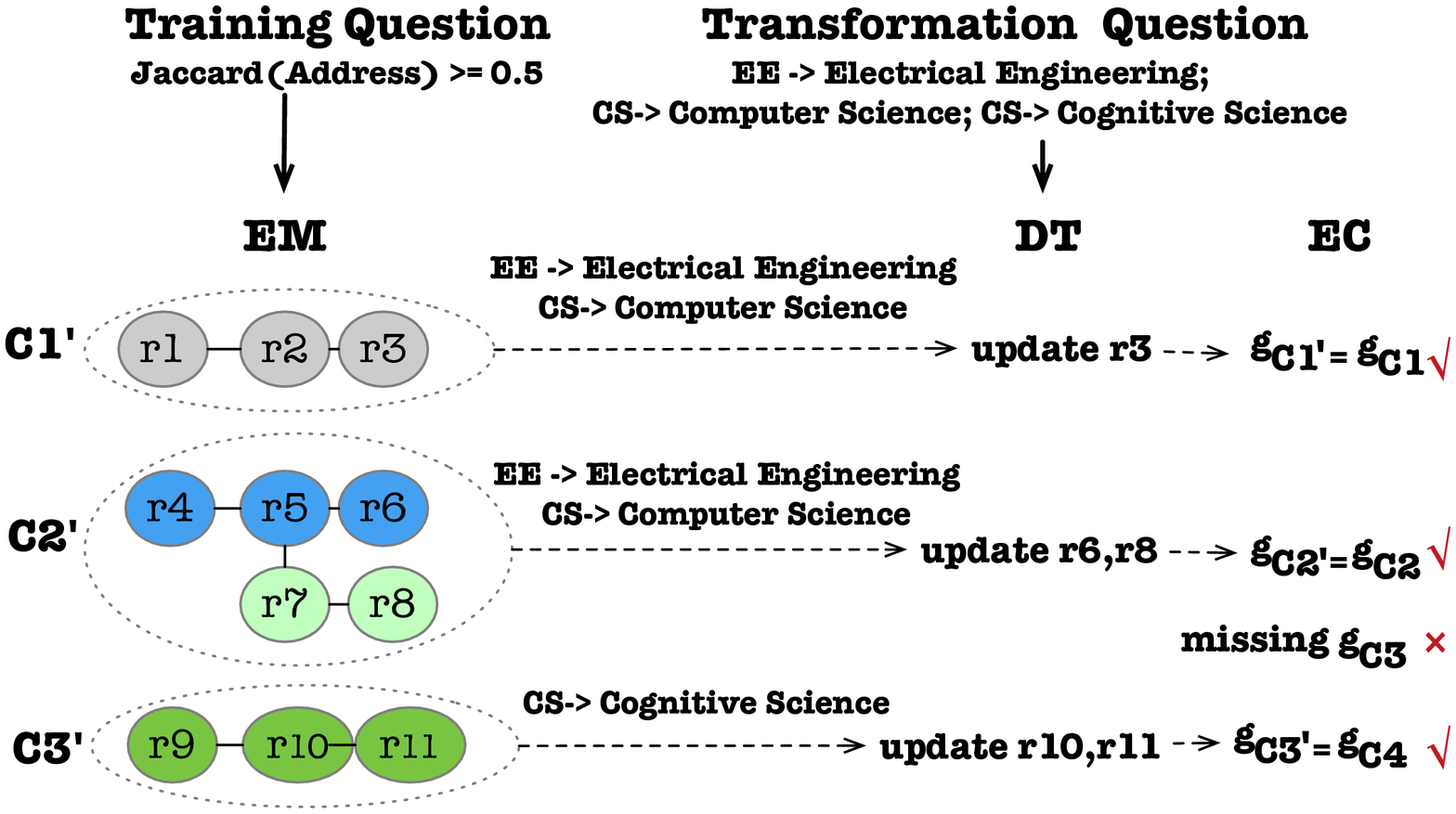}
 }

\vspace*{-1.2em}

\subfigure[{\bf DT$\rightarrow$ EM $\rightarrow$ EC}]{
     \label{fig:ex:sequential3}
     \includegraphics[width=1.05\columnwidth]{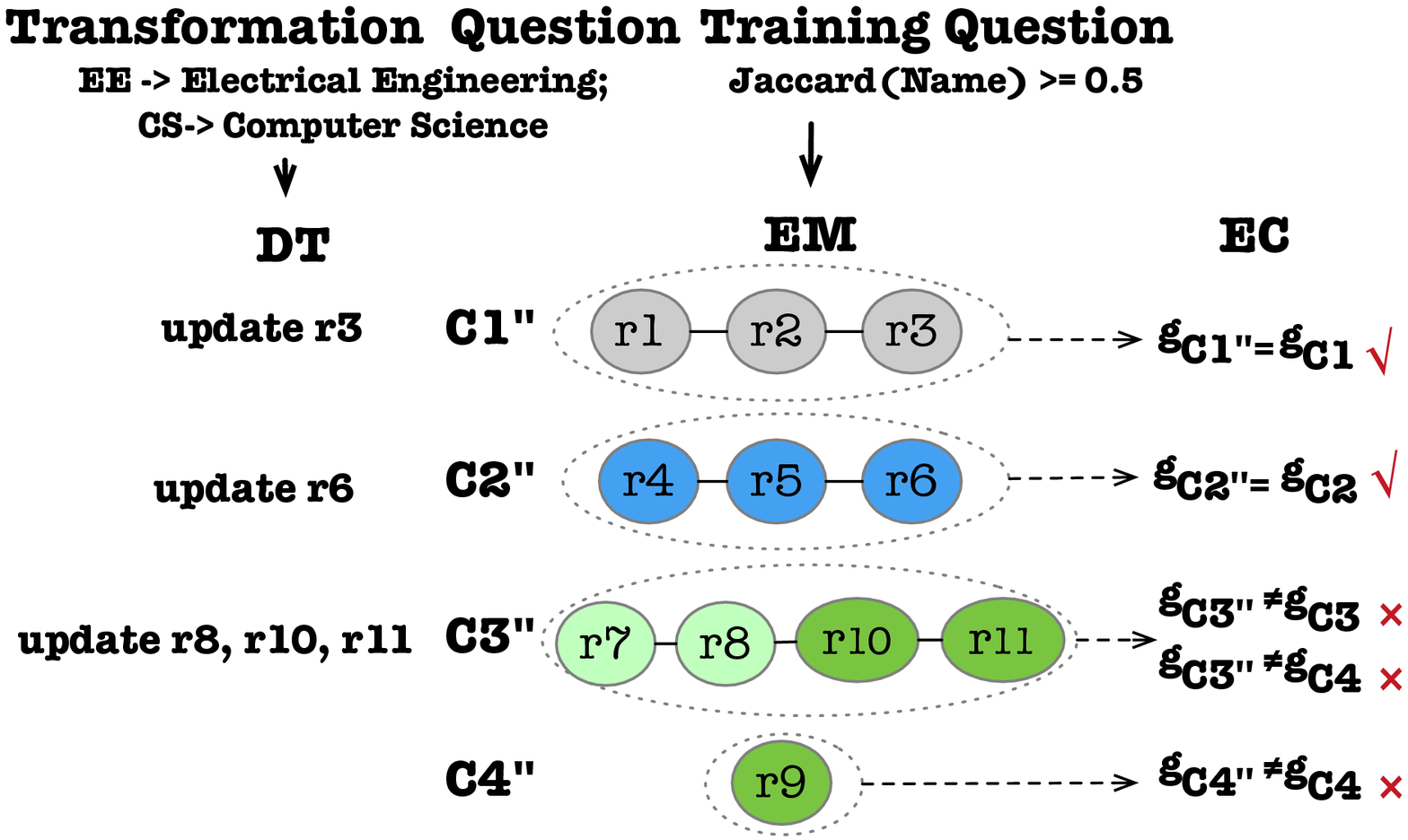}
 }
 \end{center}\vspace{-1.5em}
 \caption{Sequential Method ($r_i$: record, $g_{C_i}$: golden record of cluster $C_i$)\label{fig:ex:sequential}}
\vspace{-.5em}
\end{figure}

\begin{example}
[Shortcomings of Sequential Approaches.] 
In practice, different projects typically apply these tasks sequentially, but might be in different orders.


\stitle{(i)} {\em Entity Matching (EM) $\ra$ Data Transformation (DT) $\ra$ Entity Consolidation (EC)}
This is shown in Figure~\ref{fig:ex:sequential1}.

\be
	\item [{\bf EM}:] We first ask one training rule question (\ie~if {\it Jaccard(Address) $\geq 0.5$ then matching}), take the record pairs that obey this rule as training data, and train an EM model. Based on the EM model, we get three clusters $\{C_1', C_2', C_3'\}$. (For ease of presentation, EM is shown here using simple matching rules. However, in practice, and in this paper, the EM methods are machine learning based.) Although records \{$r_4,r_5,r_6$\} and records \{$r_7,r_8$\} refer to different real-world entities, EM incorrectly clusters them together. 
	\item [{\bf DT}:] We then ask three transformation questions, such as {\it ``EE$\ra$ Electrical Engineering?}'', and update the records. 
	\item [{\bf EC}:] After the above transformation steps, EC produces three golden records for the three generated clusters in Figure~\ref{fig:ex:sequential1}. Unfortunately, it misses the golden record for a real-world entity, $g_{C_3}$ in Table~\ref{tbl:grecords}.
\ee


\stitle{(ii)} 
{\em DT $\ra$ EM $\ra$  EC}
(Figure~\ref{fig:ex:sequential3}).

\be
	\item [{\bf DT}:] 
	We first ask two transformation questions, and transform \texttt{CS} into \texttt{Computer Science} in records $r_3,r_6,r_8,r_{10},r_{11}$. Note that \texttt{CS} should be transformed into \texttt{Computer Science} (in clusters C1 and C3) and into \texttt{Cognitive Science} (in cluster C4). However, without the clusters, it is hard to select the appropriate transformations to apply to the records. Thus \texttt{CS} in records \{$r_{10},r_{11}$\} is incorrectly transformed to \texttt{Computer Science}. 
	\item [{\bf EM}:]
	We then ask a training question and get four clusters $\{C_1'', C_2'', C_3'', C_4''\}$. 
	\item [{\bf EC}:] 
	As $r_7,r_8,r_{10},r_{11}$ are incorrectly grouped into the same cluster ${C_3''}$, EC will generate an incorrect golden record $g_{C_3''}$ that is neither $g_{C_3}$ nor $g_{C_4}$. Also, the golden record $g_{C_4''}$ is also different from $g_{C_4}$ where the Addresses are different.
\ee \vspace{-3ex}
\end{example}

\begin{figure}[!t]\vspace{-1em}
\begin{center}
\centering
\hspace*{-2em}
\includegraphics[width=0.55\textwidth]{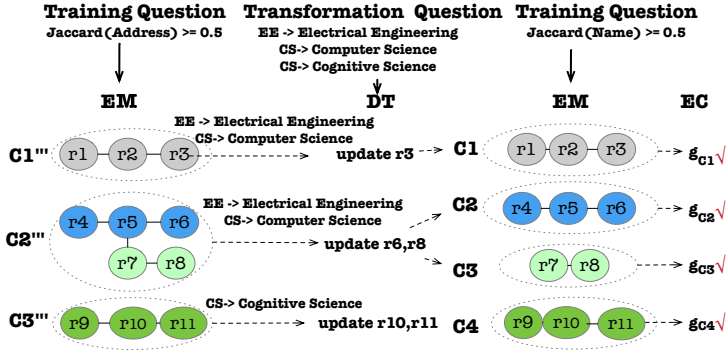}
\vspace{-1.75em}
\caption{Interleaving Questions\label{fig:ex:1}}
\end{center}
\vspace{-1.5em}
\end{figure}


Next we show the benefit of interleaving questions.

\begin{example}
[Opportunities for Interleaving Questions] We illustrate this case in Figure~\ref{fig:ex:1}.  

\be
	\item [{\bf EM}:] We first ask one training rule question and EM generates three clusters
	$\{C_1''', C_2''', C_3'''\}$. 
	\item [{\bf DT}:]Based on the cluster $C_1'''$,  we ask two transformation questions: ``\texttt{EE} $\rightarrow$ \texttt{Electrical Engineering}'' and  ``\textsf{CS} $\rightarrow$ \textsf{Computer Science}''. 
	\item [{\bf EM}:] We then ask an EM question again. Based on these transformations, the cluster \{$r_4,r_5,r_6,r_7, r_8$\} will be split into two clusters $\{r_4,r_5,r_6\}$ and \{$r_7, r_8$\}, which refer to $C_2$ and $C_3$ in Table~\ref{tbl:table}, respectively. 
	\item [{\bf EC}:] Because EM  produces correct clusters, EC is able to generate all correct golden records. 
\ee
\vspace{-3ex}
\end{example}



The above examples show that traditional methods of executing different phases sequentially are not optimal  for human involvement, and there is a need to investigate the problem of optimizing human involvement in a holistic manner. 



\vspace{-.5em}
\subsection{Research Challenges} 
\vspace{-.25em}

There are multiple types of human involvement in terms of the questions that can be asked 
and two key challenges that we need to address to holistically schedule these questions:

\hi{(1)~How to measure the benefit and cost of questions?} 
It is hard to quantify the ``benefit'' of different questions \wrt golden-record quality, because  $(i)$~we do not know the answer to each question beforehand, and $(ii)$~we have no ground-truth for golden records to be able to compute the improvement from asking a question. Moreover, the questions of the different types are not comparable because they have different optimization goals, e.g., training rule and cluster questions aim to improve the entity-matching quality while transformation questions focus on transforming variant values into the same format. Finally, different questions take different amounts of human time and we need to rank them by considering both their benefit and time cost. 

\hi{(2)~How to select ``high-quality'' questions?}  It is already expensive to estimate the benefit from and the cost of asking a question. Since there are many possible questions, it is rather expensive to enumerate all possible questions, compute the benefit and cost, and select the best one.  Moreover, questions may be correlated, and it is prohibitively expensive to enumerate all combinations. 

\vspace{-.5em}
\subsection{Contributions}  
\vspace{-.25em}

\noindent (1) We develop a human-in-the-loop framework that interleaves different types of questions to optimize the quality of golden records. We propose a question scheduling framework that judiciously selects the questions within a human time budget to maximize the accuracy of golden records (Section~\ref{sec:preliminary}).

\noindent (2) We devise cost models to measure the human time for answering different types of questions (Section~\ref{sec:preliminary}).

\noindent (3)  We propose the global benefit models to measure the quality improvement from asking different types of questions. We propose the local benefit models to greedily prune the space of possible interleaving and study the trade-off this optimization presents (Section~\ref{sec:b:1}).

\noindent (4)  We design a correlation-aware question selection method that considers correlations in selecting high-quality questions (Section~\ref{sec:b:2}).

\noindent (5)  We perform extensive experiments on three real-world datasets and show that our method significantly outperforms existing solutions on golden-record accuracy (Section~\ref{sec:exp}).

\vspace{-.25em}
\section{Holistic Data Integration}
\label{sec:preliminary}
\vspace{-.25em}

In this section, we first introduce preliminaries (Section~\ref{subsec:pre}).  We then formally define three types of human questions (Section~\ref{subsec:human}). Next we give an overview of our holistic data integration framework (Section~\ref{subsec:hybrid}).   Finally, we present a cost model to measure the human time for answering a question that we have obtained through a user study (Section~\ref{subsec:cost}).

\vspace{-.25em}
\subsection{Preliminaries} 
\label{subsec:pre}
\vspace{-.25em}

Consider a set of tables from multiple sources $\{D_1, D_2, \ldots, D_m\}$ for which schema matching has already been performed. That is, these $m$ tables contain entities from the same domain with aligned attributes. Let $\mathbb{D}$ denote the union of these tables. Our goal is to find a set of clusters of duplicate records from $\mathbb{D}$ and compute for each cluster a canonical record (a.k.a, golden record).


\vspace{-.25em}
\begin{definition} \textsc{(Golden Record)} 
Given a table $\mathbb{D}$, the {\em golden record (GR)} problem is to (1) find a set of clusters of duplicate records and (2) compute for each cluster a golden record. 
\end{definition}
\vspace{-.25em}

The golden record is typically obtained by finding clusters (\ie entity matching), transforming the variant attribute values with different formats into the same format (\ie data transformation), and merging them into canonical representations (\ie entity consolidation).



\hi{Entity Matching (EM).}
EM models decide whether two records refer to the same real-world entity, \aka a {\em match}. In this work, assume that EM is performed via an ML-based EM algorithm (for our experiments, we use random forest classifiers as they have been shown to work well in practice~\cite{DBLP:journals/pvldb/KondaDCDABLPZNP16}). Then the matching records will be grouped into the same cluster (e.g., based on transitivity or clustering algorithms).

\hi{Data Transformation (DT).}  The records may have variant values, and we use transformation rules~\cite{DBLP:journals/corr/abs-1709-10436} to transform the variant  values into the same format, \eg transforming \texttt{CS} to \texttt{Computer Science}. 



 \hi{Entity Consolidation (EC).}  Given a cluster, EC computes a canonical record for the cluster, for example, using majority voting or truth discovery based on source reliability estimation to resolve conflicts~\cite{DBLP:journals/pvldb/DongBS09a,dong2012less,
 yin2008truth,DBLP:journals/vldb/BenjellounGMSWW09,
 DBLP:conf/sigmod/RekatsinasJGPR17,Li:2014:RCH:2588555.2610509}).




\begin{table}[!t]
\vspace{-1.5em}
\centering
\caption{Notation \label{table:notation}}
{\small
\begin{tabular}{|l|l|}\hline
Notation & Description \\ \hline
$\qt(\Qt)$ & A (set of) training rule question \\\hline
$\qc(\Qc)$ & A (set of) cluster question \\\hline
$\qr(\Qr)$ & A (set of) transformation question\\\hline
\Qall & $\Qall = \Qt\cup \Qc \cup \Qr$ \\\hline
\Qs & $\Qs\subset \Qall$: selected questions \\\hline
$\budgetq(\qex)$ & Benefit from asking question $\qex$ \\\hline
$\costq(\qex)$ & Cost of asking question $\qex$ \\\hline
\end{tabular}
}
\end{table}

\begin{definition}[Accuracy of Golden Records]
The accuracy of golden records is the fraction of records whose golden records are correctly inferred among all records.
\end{definition}

\begin{example} 
[Entity Matching] 
Assume that EM takes four matching pairs as training data: $\{r_1,r_3\}$, $\{r_4, r_6\}$,  $\{r_7,r_8\}$, $\{r_9,r_{10}\}$. EM trains a model and produces 3 clusters: $\{r_1,r_2,r_3\}, \{r_4, r_5,r_6, r_7, r_8\}, \{r_9, r_{10}, r_{11}\}$.

\noindent
[Data Transformation] It transforms \text{EE and CS} to  \texttt{Electrical Engineering and Computer Science}.

\noindent
[Entity Consolidation] Considering cluster $\{r_1,r_2,r_3\}$, EC produces the golden record as $g_{C1}$ in Table~\ref{tbl:grecords}.

\noindent
[Golden Record]
Table~\ref{tbl:table} shows a table $\dataD$ with 11 records. There are four clusters (highlighted in different colors). Table~\ref{gr_example} shows the golden record for each cluster.

\noindent
$[$ Quality of Golden Record $]$ We use precision and recall to evaluate the GR quality. In Figure~\ref{fig:ex:sequential1}(or \ref{fig:ex:sequential3}), the GR precision is 1(or $\frac{2}{4}$) and the GR recall is $\frac{3}{4}$ (or $\frac{2}{4}$).
\end{example}

\vspace{-.5em}
\subsection{Human Operations} \label{subsec:human}
\vspace{-.25em}

Both EM and EC require considerable human involvement to achieve high quality results. In this paper, we consider three types of human questions. 


\vspace{-.25em}
\subsection*{\textit{2.2.1 Training Rule Questions}}
\vspace{-.25em}

There are two ways to get labeled data to train an EM model:
(i)~ask a human to validate a record pair,
or (ii)~ask a human to validate a training rule. 
For example, ``{\it if Jaccard(\texttt{Name})$\ge 0.8$ then match}'' is a training rule, and there are five pairs $\{r_1,r_2\}$, $\{r_4, r_5\}$, $\{r_8, r_{10}\}$, $\{r_{8}, r_{11}\}$, $\{r_{10}, r_{11}\}$ that obey the rule in Table~\ref{tbl:table}. For ease of presentation, we take the record pair as a special training rule which only contains one pair.


\hi{Training Rule Questions.} Formally, a matching (non-matching) rule question $\qt$ is an ``if-condition-then-match (non-match)'' clause.  Given a rule, the human is asked to approve or reject the rule.   For example, ``{\it if Jaccard$(\texttt{Name})\ge 0.5$ then match}'' is a matching rule and ``{\it if Jaccard$(\texttt{Address})\le 0.1$ then do not match}'' is a non-matching rule. To help a human better understand a training rule, we also associate a sample of record pairs  that satisfy the rule (\eg 10 pairs). We use two methods to obtain the samples: random sampling and stratified sampling. The former randomly selects pairs while the latter selects pairs from each threshold range proportionally, e.g., (0.5,0.6],  (0.6, 0.7], (0.7, 0.8], (0.8, 0.9], (0.9,1.0], based on the number of pairs in each range. We use \qt to denote a training rule and \Qt to denote a set of training rules. 


\hi{Applying a Training Rule.} If a training rule is approved by a human, the record pairs that satisfy the rule are included as training data and  the EM model is retrained; otherwise we skip the rule.

\hi{Training Rule Generation.} Training rules can be generated by humans or algorithms\cite{DBLP:journals/pvldb/WangLYF11,DBLP:journals/pvldb/MollZPMSG17} (the training pairs can be obtained by active learning). In either case, rules may be wrong,  so checking them against the training data is very important to obtain high quality training data.



\begin{example} Table~\ref{fig:tr:ex} shows 3 training rule examples. The rule ``{\it if the same \texttt{Zipcode} then match}'' will take all the records with the same \texttt{Zipcode} as matching pairs.
\end{example}

\begin{table}[!t]
\vspace{-1.5em}
\begin{center}
\centering
\hspace*{-2em}
\caption{Training Rule Questions\label{fig:tr:ex}}
\vspace{-1em}
{\small 
\begin{tabular}{|p {12.8em} | p {9.1em} | p {3.85em} |}\hline
Training Rule & Examples & Human Feedback \\ \hline
If $Jaccard(\texttt{Name})$$\ge$$ 0.5$ match & $(r_1,r_2), (r_4, r_5)$ & Yes\\\hline
If $Jaccard(\texttt{Address})$$\ge $$0.5$ match & $(r_1,r_2), (r_3, r_4)$, $(r_5, r_6)$& Yes\\\hline
If same $\texttt{Zipcode}$ match & $(r_1,r_2), (r_3, r_4),$ $(r_7, r_8)$& Yes\\\hline
\end{tabular}
} 
\end{center}
\begin{center}
\centering
\caption{Cluster Questions\label{fig:cl:ex}}
\vspace{-1em}
{\small 
\begin{tabular}{|l | l |}\hline
Cluster & Human Feedback \\ \hline
\{$r_4,r_5,r_6,r_7,r_8$\} & No: $\{r_4,r_5,r_6\}$; \{$r_7,r_8$\}\\\hline
\{$r_4,r_5,r_6$\} & Yes\\\hline
$\{r_1,r_2,r_3\}$ & Yes \\\hline
\end{tabular}
} 
\end{center}
\begin{center}
\centering
\hspace*{-1.5em}
\caption{Transformation Rule Questions\label{fig:gr-example}}
\vspace{-1em}
{\small 
\begin{tabular}{|l|l|l|}\hline
Transformation Rule & Examples & Human Feedback  \\ \hline
\#th $\rightarrow$ \# (\# is a numerical value) & $r_3,r_5$ & Yes\\\hline
EE $\rightarrow$ Electrical Engineering & $r_3,r_6$ & Yes\\\hline
CS $\rightarrow$ Computer Science & $r_3,r_8$ & Yes\\\hline
\end{tabular}
} 
\vspace{-.5em}
\end{center}
\end{table}

\vspace{-.25em}

\subsection*{\textit{2.2.2. Cluster Questions}}
\vspace{-.25em}

The EM model may generate incorrect clusters. 
In practice, humans must also be involved in verifying clusters.

\hi{Cluster Validation Questions.}  Formally, a cluster question $\qc$ is a cluster of records that asks a human to verify whether the records in the cluster refer to the same entity. If they do, the human approves the cluster; otherwise the human is required to split the cluster into $C_1, C_2,\cdots, C_y$ sub-clusters, such that records in each sub-cluster represent the same entity.  We use \qc to denote a cluster question and \Qc to denote a set of cluster questions.

\hi{Applying A Cluster Question.}  
If a cluster is approved, each pair of records within the cluster is treated as a matching pair that can be used to enhance the EM model; otherwise, the pairs in the sub-clusters, obtained after the split, are treated as matching pairs and all pairs from different sub-clusters are treated as non-matching pairs. These pairs are used as training data to enhance the EM model.


\hi{Cluster Question Generation.}  All the clusters generated by the EM model can be used as cluster questions. In practice, the cluster is not large, usually about 10 records. We discuss how to support large clusters in Appendix~\ref{app:large}.

\begin{example} Table~\ref{fig:cl:ex} shows 3 cluster questions.  Consider cluster \{$r_4,r_5,r_6,r_7,r_8$\} in Figure~\ref{fig:ex:1}. 
A human discriminates the address of \texttt{Harvard Electrical Engineering} from \texttt{Harvard Computer Science}, and splits the cluster into two sub-clusters: $\{r_4,r_5,r_6\}$ and $\{r_7,r_8\}$.  Then the matching pairs $(r_4, r_5), (r_4, r_6), (r_5,r_6), (r_7,r_8)$, and non-matching pairs $(r_4,r_7), (r_4,r_8), (r_5,r_7), (r_5,r_8), (r_6,r_7), (r_6,r_8)$ are used as training data to enhance the EM model.
\end{example}
\vspace{-.5em}

\subsection*{\textit{2.2.3 Transformation Rule Questions }}


We use transformation rules~\cite{DBLP:journals/corr/abs-1709-10436}  as a way to transform the variant data values into the same format to improve the quality of golden records.

\hi{Transformation Rule Questions.}  A transformation rule question is of the form $v\rightarrow v'$. For example, Table~\ref{fig:gr-example} shows several transformation rules. \#th $\rightarrow$ \# will transform a numerical value with \texttt{th} to the numerical value, \eg 50th is transformed to 50 in record $r_3$ and 29th is transformed to 29 in records $r_5,r_8$. We ask a human to verify whether $v$ should be transformed to $v'$.  To help the human better understand a transformation rule, we show sample records with value $v$, and the human can check these records to decide whether to apply this rule. We use \qr to denote a transformation rule and \Qr to denote a set of transformation rules.

\hi{Applying a Transformation Rule.}  Given a rule  $v\rightarrow v'$, if the rule is approved, we transform $v$ to $v'$ for all records with value $v$, and update the table $\dataD$.

\hi{Transformation Question Generation.}  We use existing techniques to generate transformation questions~\cite{DBLP:journals/corr/abs-1709-10436}.  Here, we briefly introduce the basic idea and refer the reader to~\cite{DBLP:journals/corr/abs-1709-10436} for more details.  A simple way is to enumerate every pair $(v, v')$ of two non-identical tokens in an attribute. Then for each pair $(v, v')$, we count the number (frequency) of clusters that contain the pair. Next we select most frequent pairs as transformation rules. A more efficient way is to align the tokens and only enumerate the aligned pairs. For example, first split the attribute values into a sequence of tokens, then compute the longest common subsequence (LCS), and use the LCS to align the tokens.


\begin{example}
Consider the \texttt{Address} attribute for cluster $r_1,r_2,r_3$. The aligned token sequences are 

``\texttt{50$~~~~~~~~$ $~~~~$| Vassar | St | Cambridge | MA}''

``\texttt{50$~~~~~~~~$$~~~~$ | Vassar | St | Cambridge | MA}''

``\texttt{50th | Vassar | St | Cambridge | MA}''

Then $(50th, 50)$ is an aligned pair and $50th \rightarrow 50$ is a possible transformation rule. Similarly  $(29th, 29)$ is an aligned pair. The two pairs can be merged by a regular expression~\cite{DBLP:journals/corr/abs-1709-10436} and a generalized transformation rule is $\#th \rightarrow \#$. 
\end{example}

\vspace{-1em}


\subsection{Optimizing Human Involvement} \label{subsec:hybrid}

\begin{figure}[t!]\vspace{-.5em}
\begin{center}
\centering
\hspace*{-1.5em}\includegraphics[width=0.56\textwidth]{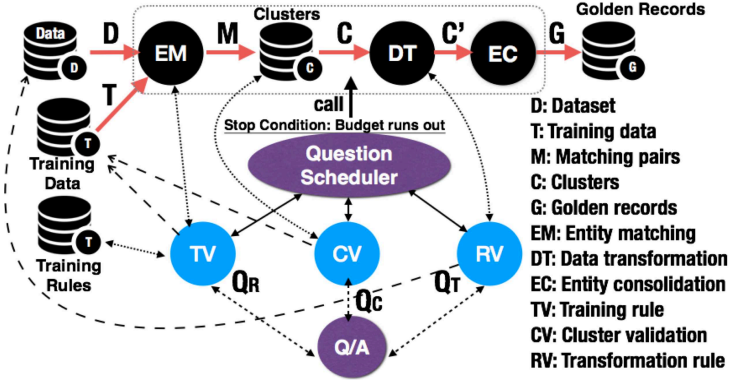}
\vspace{-2em}
\caption{Architecture of Holistic Data Integration\label{fig:arch1}}
\end{center}
\vspace{-2em}
\end{figure}

Evidently, there will be a large number of training/cluster/transformation questions, and we cannot ask all of them. 
Instead, we propose a human-in-the-loop framework that judiciously selects the most beneficial questions to ask. Figure~\ref{fig:arch1} shows the workflow. The key point is that, different questions may be scheduled in an arbitrary order, and the pipeline will be rerun as  questions are answered. 

\hi{(1) Matching Algorithms.}  We first train an EM model and run the model on table $\dataD$ to generate a set of  clusters. For each cluster, if there are some transformation rules, we update the records by applying these rules, and then run an EC algorithm to generate the golden record for this cluster (e.g., using a majority voting algorithm to vote the golden value on each attribute).  Note that we need some training data to train the EM model, which we obtain through training rules. We will discuss how to select training rules in Section~\ref{subsec:local}. 

\noindent {\bf (2) Question Generation}.  
$\Qt$, $\Qc$ and $\Qr$ are generated based on the results of the generation algorithms as discussed in Section~\ref{subsec:human}.  
Let $\Qall = \Qt\cup \Qc\cup \Qr$ denote the set of all possible questions.

\noindent {\bf (3) Iterative Question Scheduling}. We iteratively select some questions from the three question sets and ask a human for feedback. For a training rule question \qt and a cluster question  \qc, the outcome will be more training data for training the EM model; for a transformation rule question, we update dataset $\dataD$. Then, based on the refined training data and updated dataset, we rerun the machine algorithms to compute the golden records, and update the three question sets $\Qt$, $\Qc$, and $\Qr$. We iteratively call this step until the budget is consumed.

\hi{Optimization Goal.}  Given a table $\dataD$ and a budget $\budget$, we aim to select a sequence of questions $\Qs=\langle q_1,q_2, \cdots, q_\budget\rangle$ in order to maximize the quality of golden records, where $q_i$ is a training, cluster, or transformation question.

A brute-force question scheduling method enumerates every $B$-size subsets of $\Qall$, computes the benefit from asking these $B$ questions that measures the golden record quality improvement (before and after asking these $B$ questions), and selects the subset with the largest benefit/cost ratio. However this method has several limitations.


\hi{Limitation 1: Question Sequence Enumeration.} It is prohibitively expensive to enumerate all size-$B$ subsets of $\Qall$. 

\hi{Limitation 2: Golden Record Quality Computation.} It is hard to compute the golden record quality of a set of questions, because $(i)$ we do not know the answer of each question beforehand, and $(ii)$ we do not know the ground truth of golden records. 

\hi{Limitation 3: Questions may be Correlated.}  Asking a question may affect the clusters and golden records, and thus affect other questions. So \Qall will dynamically change after asking some questions. We do not want to select a static set of questions. Instead, we first select several questions, use human feedback on these questions to compute the cluster quality and golden record quality, and then utilize the results to guide the selection of the subsequent questions. 

To address the above limitations, we propose an iterative estimation based method. The basic idea is to select $b$ questions in each round, use the human feedback on these questions to estimate the benefit of each question (and a set of questions), and then utilize the benefits to select $b$ questions in the next round. (We will discuss how to choose an appropriate $b$ later.) Through multiple iterations, this method can adaptively select the most beneficial questions to ask.

\hi{Question Scheduling Framework.}  Algorithm~\ref{algorithm:framework} shows the pseudo code of our question scheduling framework. 

\hi{1. Initial Step.}  It first trains an EM model and generates a set of questions \Qall (line 1).

\hi{2. Benefit Inference.}   It estimates the benefit and cost of questions in \Qall (line 3). We will discuss how to compute the benefit and cost later. 

\hi{3. Question Selection.} It selects $b$ questions \Qb (line 4).

\hi{4. Machine Algorithm.} It asks the questions in \Qb, runs the machine algorithms to compute the golden records, and updates the question set \Qall (lines 5-6).


\begin{figure}[!t]
\begin{algorithm}[H]
\caption{QuestionScheduler\label{algorithm:framework}}
\linesnumbered \SetVline
\KwIn{Table $D$, A Training Rule Set \Qt}
\KwOut{Golden-record set $G$ of $D$}
ColdStart()\;
\While{$B>0$}
{
	BenefitInference(\Qall)\;
	\Qb= QuestionSelection(\Qall)\; 
	Ask questions in \Qb\;
	\Qall = MachineAlgo(\Qb)\;
	$B=B-b$\;
} 
\end{algorithm}

\begin{function}[H]
\caption{MachineAlgo()}
\linesnumbered \SetVline
\KwIn{Table $D$, \Qb}
\KwOut{\Qt, \Qc, \Qr}
Train/update the EM model based on the answer of \Qb\;
Compute clusters based on the EM model\;
Produce golden-records using EC algorithms on clusters\;
Compute \Qt, \Qc and \Qr\;
\end{function}
\vspace{-1.5em}
\end{figure}

\hi{Discussion.} Obviously when $b$ is small, this algorithm can enumerate all the $b$-size subsets; but it may neglect the correlations for a larger question pool. If $b$ is large, it is impossible to enumerate all $b$-size subsets, because it requires to consider too many subsets and calls the machine algorithms many times. To address this issue, we first consider a simple case $b=1$ in Section~\ref{sec:b:1}, then discuss how to support the case of $b\ge 2$ in Section~\ref{sec:b:2}.

\subsection{Cost Model for Human Feedback} 
 \label{subsec:cost}


Because our proposed framework is centered around obtaining human feedback, we need a way to estimate the cost of such involvement.  
As different questions take different human time (called cost),  it is important to measure the cost of answering a question. We first qualitatively compare the cost of different questions and then present a quantitative model to measure the cost based on a user study.

\hi{Training Rule Question.} 
Suppose each rule \qt contains $|\qt|$ (e.g., 10) record pairs and it takes a unit time for a human to check a pair, leading to a human time of $|\qt|$ to check all the pairs. However, most of the pairs are similar as they satisfy the same rule, and the human does not need to examine the pairs one by one. Instead she can check multiple pairs together, and thus the cost of a training rule question $\qt$ is smaller than $|\qt|$.

\hi{Cluster Validation Question.} A cluster question may cover many records, and a human is required to check all the records and split them into different sub-clusters. In the worst case, the cluster is split into many sub-clusters and the cost is quadratic to the number of records in \qc (i.e., $|\qc|^2$). In the best case, the cluster is not split and the cost is linear to $|\qc|$.  

\hi{Transformation Question.} Given a transformation rule question, a human checks the records associated with the question and decides whether to apply the transformation to the records. Since most records that obey the rule belong to different clusters and are not similar, the human requires to check the records one by one. Thus the cost is linear to the number of records associated with \qt (i.e., $|\qc|$).

\hi{Cost Model.}  Based on the above observations, we propose a cost model to quantify the human cost of answering a question, where the cost is a function of the question size. We conducted a user study to construct the cost model. 
For each question type, we varied the question sizes from 1 to 100. For each question size, we selected 100 questions, assigned each question to 10 students, recorded the time to answer the question, and computed the average time as the human cost for this question size. Next we used regression models to construct the cost model. We tried different models, e.g., linear distribution, polynomial distribution, exponential distribution, and logarithmic distribution, and selected the one that best fits the distribution of human time for answering different questions.

Based on the experimental results from our user study, we find that the cost of answering a training rule question follows the logarithmic distribution, best fit by the function:
\begin{equation}
\mathcal{C}(\qt)=8\log_e(|\qt|+3) - 10.
\end{equation}

The cost of a cluster question follows the quadratic polynomial distribution, best fit by:
\begin{equation}
\mathcal{C}(\qc)=\frac{|\qc|^2}{100}+\frac{|\qc|+1}{5}.
\end{equation}

The cost of a transformation question follows the linear distribution, best fit by:
\begin{equation}
\mathcal{C}(\qr)=\frac{|\qr|+0.5}{1.5}.
\end{equation}

The relative error between the human time and our  cost model is smaller than 5\% in our user study, and the result is shown in Appendix~\ref{app:exp:costmodel}. Any sophisticated cost model can be integrated into our method, and we leave out the study of such models as future work.

\section{One Question Per Iteration} \label{sec:b:1}

In this section, we define a benefit model to measure the golden record quality improvement from asking a single question. 
We then select the question that has the largest benefit/cost ratio in each iteration.   We address the problem of selecting a batch
of questions at a time in Section~\ref{sec:b:2}.


\subsection{Global Benefit Model} 
\label{subsec:global}


We say a question incurs a benefit if it helps obtain more correct golden records.  
If  $x$  golden records are correct before asking the question and $y$ are correct after asking the question and rerunning the algorithm, then the benefit is
$y-x$.  
There are two challenges in computing this benefit:

\hi{$(i)$.~The answer to a question is unknown beforehand.} 
To tackle this challenge, we would need to enumerate all possible answers to the question,  compute a probability for each possible answer, and calculate the {\it expected benefit} from asking the question. 

Formally, considering a question \qex, let $\{a_1,a_2,\cdots, a_n\}$ denote the set of possible answers of \qex,  $\pro(q=a_i)$ denote the probability that $q$'s answer is $a_i$, and $\benefitq(q=a_i)$ denote the benefit from asking question \qex whose answer is $a_i$. 
The expected benefit of asking question $\qex$ can be computed as:
\begin{equation}
\benefitq(q)=\sum_{i=1}^n\pro(q=a_i)\benefitq(q=a_i).
\end{equation}

We discuss how to compute  $\benefitq(q=a_i)$ and $\pro(q=a_i)$ later.



\hi{$(ii)$.~There is no ground truth for golden records.} 
To address this issue, we compare $G$ and $G'$, the sets of golden records before and after asking question $\qex$, respectively\footnote{As different records may have the same golden records, we use a multi-set to represent the set.}.
We compute the difference $G'-G$ and call $|G'-G|$ the number golden record changes. If the new golden records in $G'-G$ are all correct,  the benefit from asking question $\qex$ is $|G'-G|$. 
Although the ground truth is unknown, we use $|G'-G|$ as an estimate of the number of correct golden records in $G'-G$ as in general asking questions will improve the quality of the golden records, and we expect humans to make few mistakes.



\subsection*{\textit{(1) Computing Benefit $\benefitq(q=a_i)$}}

Let $\nuq(\qex=a_i)=|G'-G|$ denote the number of changes in the golden records if the query answer is $a_i$. We estimate the benefit from asking question \qex \wrt  the answer $a_i$ as:
\begin{equation}
\benefitq(q=a_i)=\nuq(q=a_i).
\end{equation}


\subsection*{\textit{(2) Computing Probability $\pro(\qex=a_i)$}}


We now discuss how to compute the probability $\pro(\qex=a_i)$ for the three types of questions we introduced earlier. 

\hi{Training Rule Questions.}
There are two possible answers for a training rule question: either the rule is approved (denoted by Y) or rejected (denoted by N). Since $\pro(\qt=N)=1-\pro(\qt=Y)$, we focus on computing $\pro(\qt=Y)$. 

The EM model can return a probability of whether a pair $p$ of records is matching, denoted by $\pro(p=Y)$.
Consequently, 
we use the average probability for all pairs in \qt to compute the probability $\pro(\qt=Y)$ for the rule, \ie 
\begin{equation}
\pro(\qt=Y)=\frac{\sum_{p\in \qt} \pro(p=Y)}{\sum_{p\in \qt} 1}.
\end{equation}

%


\hi{Cluster Questions.}
The answer to a cluster question \qc depends on the answers
to the individual pairs of records. We thus need to consider all such pairs.
There are $|\qc| \choose 2$ pairs in the cluster,   denoted by $p_1,p_2,\cdots, p_{|\qc| \choose 2}$.
Each pair has only two possible answers, matching or non-matching. 
Thus, there are $2^{|\qc| \choose 2}$ possible answers $(p_1=x_1,p_2=x_2,\cdots, p_{|\qc| \choose 2}=x_{|\qc| \choose 2})$, where $x_i\in\{Y,N\}$. $\pro(p_i=Y)$ can be computed using the EM model and  $\pro(p_i=N)=1-\pro(p_i=Y)$. 
Consequently, we compute the probability of each possible answer using:
\begin{equation}
\pro\big(\qt=(p_1=x_1,\cdots, p_{|\qc| \choose 2}=x_{|\qc| \choose 2})\big)=\prod \pro(p_i=x_i).
\end{equation}

If the cluster is large, it will be prohibitively expensive to enumerate every possible case. 
To address this issue, we only consider the cases with large probabilities.  
For example, if the probability $\pro(p_i=Y)$ is large (\eg larger than 0.8), $\pro(p_i=N)$ will be small and we ignore the case of $p_i=N$, \ie ignoring all the possible answers $(p_1=x_1,p_2=x_2,\cdots, p_i=N, \cdots, p_{|\qc| \choose 2}=x_{|\qc| \choose 2})$. 

To further improve the performance, we propose a statistics-based method. Given a record pair in a cluster, the EM model computes a probability for the pair. If the probability is larger than 0.5, this pair will be taken as a matching pair (as the matching probability is larger than the non-matching probability).  However many pairs in the clusters may not represent the same entity, implying we need to split the cluster. Usually the cluster is split based on some probability threshold. For example, a pair is actually matching if its probability is larger than a threshold $\tau=0.8$. We can split the cluster into several sub-clusters based on this threshold, by building a graph for the records, where the vertices are records and there is an edge between two records if their matching probability is larger than the threshold.  The records in the same connected component then belong to the same cluster.  For each possible choice of $\tau$, we can generate a set of such sub-clusters (i.e., connected components). However, it is expensive to enumerate every threshold. 

To address this issue, we can use a fixed number of thresholds, \eg $\tau \in (0.5, 0.6, 0.7, 0.8, 0.9)$, to split a cluster, each of which will induce a set of sub-clusters from the cluster. For each such set of sub-clusters, we can compute the likelihood of that sub-cluster set using the clusters provided by humans so far. Suppose a cluster question is answered by a human and $C_h$ is the set of sub-clusters provided by the human. Our goal is to find the value of $\tau$ whose sub-cluster set $C_\tau$ best matches $C_h$. To do this, we need to compute the set similarity between $C_h$ and $C_\tau$ for each value of $\tau$; this can be done using any set similarity function, e.g., Jaccard similarity. If multiple clusters have been answered by humans, we can compute the likelihood of $\tau$ based on the percentages of clusters whose best matching thresholds are $\tau$. In this way, given a cluster question, we can generate a limited number of its answers as well as the probabilities based on the thresholds.





\hi{Transformation Questions.}
There are two possible answers for a transformation  rule question: approved (Y) or rejected (N). Since $\pro(\qr=N)=1-\pro(\qr=Y)$, we focus on computing $\pro(\qr=Y)$.
Suppose $\qr=v\rightarrow v'$, and there are $\nuq(v|\qr)$ records with value $v$ and $\nuq(v'|\qr)$ records with value $v'$. Obviously, if most of the records are with value $v'$, then the rule has a high probability to be approved. 
Thus, we can compute  $\pro(\qr=Y)$ as follows\footnote{For the aggregated rules, \eg \#th $\rightarrow$ \#, we can also compute $\nuq(v|\qr)$ and $\nuq(v'|\qr)$ by finding sets of records that respectively contain $v$ and $v'$.}:
\begin{equation}
\pro(\qr=Y)=\frac{\nuq(v'|\qr)}{\nuq(v'|\qr)+\nuq(v|\qr)}.
\end{equation}

\hi{Discussion.} It will be  prohibitively expensive to compute the global benefit if there are many possible questions, 
because this would require enumerating every possible answer for all questions and run the machine algorithm to compute the golden records. To address this problem, we introduce a more efficient method, as discussed below. 

\subsection{Local Benefit Model} 
\label{subsec:local}


As noted in the previous section, it is time consuming to compute the global benefit, because it is expensive to enumerate all possible answers and rerun the EM and EC algorithms to compute the number of changes in the golden records. To avoid rerunning the EM and EC algorithms, we can rank the questions from the same set  ({i.e., training rule, cluster, transformation}), select the top-$k$ questions from each set, and compute the global benefit of these selected $3k$ questions and choose the one with the largest global benefit/cost ratio. Thus the local benefit uses a coarse-grained way to prune away the questions that have smaller probabilities of having large global benefit. We provide the complexities of computing global benefits and local benefits in Appendix~\ref{appendix:complexity}. 

Toward this goal, we compute a ``{\it local benefit}'' that measures the importance of questions from the same set and utilize it to rank the local questions. 
 Specifically, because training rule and cluster questions aim to generate more training data in order to improve the quality of EM,  we first rank these questions based solely on how much they improve the EM quality. Then, because transformation questions aim to transform the variant values to obtain a canonical value,  we rank them by the frequency of the question, e.g., the number of records that can be applied by this transformation.  Next we give formulas for computing these local benefits.

\hi{Training Rule Questions.} In Section~\ref{subsec:global}, the benefit of a training rule question in the global setting was estimated based on its overall expected impact on the number of golden records.  In contrast, in the local context, 
the goal is to simply choose training rule questions that ask users to verify the highest-value training examples.
 The value of a training example depends on several factors, including its coverage,  (\ie how many record pairs satisfy the rule),  accuracy (\ie how many pairs satisfying the rule are correct), and utility (\ie whether it actually improves the EM model).
 For example, given a rule ``{\it if the same \texttt{Zipcode} then match}'', if there are 16 pairs satisfying the rule and 10 pairs are correct; the coverage is 16 and the accuracy is $\frac{10}{16}$. 
Its utility depends on whether the matched (or non-matched) records are  easily discriminated by the EM model.
Thus, we aim to select questions that verify training rules with high coverage, high accuracy, and containing many high utility pairs. 
We discuss next how to compute the coverage of a training rule, the accuracy of a training rule, and the utility of a record pair. 

\hi{\it (1) Computing the coverage $\cov(\qt)$ of a training rule \qt.} A straightforward method would enumerate all the record pairs and identify the pairs that obey the rule \qt.  
This method is not scalable for large datasets. To address this issue, we can use similarity join algorithms~\cite{DBLP:journals/pvldb/JiangLFL14}, which first generate signatures for each record, then take the pairs of records that share common signatures as candidate pairs, and finally verify the candidate pairs by checking whether they actually obey the rule. Since these algorithms can use signatures to prune many dissimilar pairs, they scale well~\cite{DBLP:journals/pvldb/JiangLFL14}.


\hi{\it (2) Computing the accuracy $\pre(\qt)$ of \qt.}  If the training rule is written by an expert, we can ask the expert to provide a confidence. If the training rule is generated by algorithms, the algorithms also provide a confidence~\cite{DBLP:journals/pvldb/WangLYF11,DBLP:journals/pvldb/MollZPMSG17}. Then we can take the confidence as the probability.



\hi{\it (3) Computing the utility $U(\qt)$ of \qt.}  We first  use the EM model to compute a probability $Pr(p)$ that the pair $p$ is a matching pair. The larger $Pr(p)$ is, the most likely $p$ is a matching pair.  The smaller $Pr(p)$ is, the most likely $p$ is not a matching pair. Note that if $Pr(p)$ is close to 0.5, the EM model cannot discriminate the pair. We want to ask the human to label such ``uncertain'' pairs and use the answer as training data to enhance the EM model. 
To this end, we define the entropy of a pair $U(p)$ as below.
\begin{equation}\vspace{-.5em}
U(p)=-\Bigl(\log Pr(p) + \log\bigl(1-Pr(p)\bigr)\Bigr)
\end{equation}
The larger the entropy is, the smaller the utility is. So we compute the utility by normalizing the entropy as below 
\begin{equation}
\uti(p)=1-\frac{U(p)}{MaxU}
\end{equation}
where $MaxU$ is the maximal entropy among all pairs.  

Based on the three factors, we compute a local ranking score for a training rule,
\begin{align}
S(\qt)&=\pre(\qt)\cov(\qt)\frac{\sum_{p\in \qt}\uti(p)}{\sum_{p\in \qt}}\\
&=\pre(\qt)\sum_{p\in \qt}\uti(p)
\end{align}
We rank the rules by score/cost ratio, \eg $\frac{S(\qt)}{\costq(\qt)}$, in descending order and select the top-$k$ questions following this order.



\hi{Cluster Questions.} As with training-rule questions, in the local context our goal is to choose clusters for humans to verify that most improve the EM model, without considering the global impact on the number of golden records.
If all pairs in a cluster refer to the same entity, then the cluster has a low benefit to improve the EM model. However, if most pairs are hard to be discriminated by the EM model, then based on the human feedback, we can improve the EM model. Thus, we want to use the utility of record pairs in the cluster to quantify the cluster question. Thus, we use the sum of the utility of these pairs to compute a score of a cluster as below:
\begin{equation}
S(\qc)=\sum_{p\in \qc \times \qc }\uti(p)
\end{equation}
We sort the clusters by the score/cost ratio in descending order and select the top-$k$ questions following this order. 


\hi{Transformation Questions.} For transformation questions, we compute their local benefit based simply on how many records they can be used to transform.  Specifically, 
each transformation question verifies a transformation rule $\qr=v \rightarrow v'$, which applies to a specific $v$. Let $|\qr|$ dente the frequency of \qr, i.e., the number of clusters that contain the pair $(v,v')$.  The more frequent  $|\qr|$ is, the more transformations can be applied.  Thus we use the rule frequency to compute a score for a transformation rule as below:
\begin{equation}
S(\qr)=|\qr|
\end{equation}
We rank the questions by the score/cost ratio in descending order and select the top-$k$ questions following this order.



\subsection{Selecting $k$}

%

Obviously, there is a tradeoff between efficiency and quality to select an appropriate $k$. 
A small $k$ leads to high efficiency but low accuracy of golden records, because the most beneficial questions may not be in local top-$k$ questions. 
A larger $k$ leads to high quality but low efficiency,  because it needs to enumerate many more questions to compute the global benefit.  
As a tradeoff, we first set $k=B$ and then tune $k$ in each iteration based on the global benefits of the selected questions as follows. For each question type, suppose $q$ is the ``most  beneficial'' question in this  type that has the largest global benefit/cost ratio and its local ranking position is $k'$. In each iteration we compute $k'$ and use $k'$ as an estimation of $k$ for the next iteration.

\section{Multiple~Questions~Per~Iteration}
\label{sec:b:2}

In this section, we address the case where multiple questions are answered in each iteration. 
We fist consider the case of $b=2$ (Section~\ref{subsec:b2}) and then generalize our techniques to support $b>2$ (Section~\ref{sec:b:x}).  We then consider how to select an appropriate value of $b$ (Section~\ref{sec:b:set}).

\subsection{Question Selection for {\large $b=2$}}\label{subsec:b2}

Selecting the two questions with the largest benefit might be worse than selecting two highly correlated questions, because the correlated questions may have mutual positive influence on each other. Thus, we propose a correlation-aware question selection method. 



Let us first discuss how to compute the global benefit  $\benefitq(q,q')$ from asking two questions $q$ and $q'$.  Let $\pro(q=a_i,q'=a'_j)$ be the probability that the answer of $q$ is $a_i$ and the answer of $q'$ is $a'_j$, and $\benefitq(q=a_i,q'=a'_j)$ be the benefit of asking questions $q,q'$ together with answers $a_i,a'_j$. 
The global benefit  $\benefitq(q,q')$ is computed as follows:
\begin{equation}
\benefitq(q,q')=\sum_{i=1}^n\sum_{j=1}^m\pro(q=a_i,q'=a'_j)\benefitq(q=a_i,q'=a'_j).
\end{equation}

We assume that the answers to two questions are independent, and we compute $\pro(q=a_i,q'=a'_j)$ as below.
\begin{equation}
\pro(q=a_i,q'=a'_j)=\pro(q=a_i)\pro(q'=a'_j).
\end{equation}

Let $\nuq(\qex=a_i, q'=a'_j)$ be the number of changes on the golden records from asking questions $q,q'$.  We estimate the benefit from asking questions $\qex,q'$ with the answers $a_i,a'_j$ as below:
\begin{equation}
\benefitq(q=a_i,q'=a'_j)=\nuq(q=a_i,q'=a'_j). 
\end{equation}

The  correlation-aware question scheduling algorithm iteratively calls the following steps until the budget is consumed.  


\hi{1. Correlation-aware Benefit Inference.}  We first identify the top-$k$ questions with the largest local benefits from each question set. We then enumerate each question pair and compute its global benefit.

\hi{2. Correlation-aware Question Selection.} We select the question pair with the largest global benefit, ask the two questions, and rerun the machine algorithms.

\subsection{Question Selection for {\large $b>2$}}\label{sec:b:x}

When $b>2$, we can still use the above algorithm for $b=2$ where 
we select $b$ questions with the largest benefit in each iteration. 
Thus, we enumerate all $b$-size subsets of the selected $3k$ questions (note that we select the top-$k$ questions from each type), and compute the benefit of each subset. However, this method has two limitations. First, it needs to enumerate  $3k \choose b$ cases and is rather expensive when $k$ or $b$ are large. Second, it is expensive to estimate the benefit from asking $b$ questions together as it needs to enumerate the permutations of all possible answers of the $b$ questions. 


We propose two techniques to alleviate these limitations. First, we partition the $3k$ questions into multiple groups such that (1)~the questions in different groups have no correlation and (2)~the questions in the same group have correlations. Hence, we can avoid considering the question correlations from different groups. Second, we use the benefit of two questions to estimate the benefit of $b$ questions.


\hi{Question Grouping.} We first define whether two questions are correlated and then partition the questions that have no correlations into different groups.

\begin{definition}\textsc{(Correlation)}
Two questions $q,q'$ are  positively correlated if $\benefitq(q,q')>\benefitq(q)+\benefitq(q').$
Two questions $q,q'$ are negatively correlated if $\benefitq(q,q')<\benefitq(q)+\benefitq(q').$
\end{definition}
\vspace{-.5em}
\begin{definition}
\textsc{(No Correlation)}
Two questions $q,q'$ are not correlated if $\benefitq(q,q')=\benefitq(q)+\benefitq(q').$
\end{definition}

We enumerate every question pair and compute the benefit. Then, we simply put all question pairs with correlations in the same group, using the definitions of correlations above. This process generates a set of disjoint groups $P_1, P_2, \cdots, P_{|P|}$.

\hi{Benefit Estimation from Asking a Set \Qb of $b$ Questions.} Let $\Qb_i=\Qb\cap P_i$. The questions in \Qb are split into $|P|$ groups $\Qb_1, \Qb_2, \cdots, \Qb_{|P|}$ such that the questions in the same group have correlations and the questions from different groups have no correlations. Thus, we can compute the benefit of \Qb as below: 
\begin{equation}\label{eq:global:sum}
\benefitq(\Qb)=\sum_{i=1}^{|P|} \benefitq(\Qb_i)                                                                                                                                                                                                                             
\end{equation}

If $|\Qb_i|$ is large, it is still expensive to compute $\benefitq(\Qb_i)$. To address this issue, we propose an approximate method to estimate $\benefitq(\Qb_i)$. The basic idea is that we use the average pairwise correlation to  estimate the overall correlation. Let $ \frac{\benefitq(q',q'')}{\benefitq(q')+\benefitq(q'')}$ denote the correlation between $q'\neq q''\in \Qb_i$. We use the average pairwise correlations  to estimate the correlation among multiple questions in $\Qb_i$, \ie  $\frac{\sum_{q'\neq q''\in \Qb_i} \frac{\benefitq(q',q'')}{\benefitq(q')+\benefitq(q'')}}{|{\Qb_i \choose 2}|}$. Then we can compute the benefit from asking questions in $\Qb_i$ as below.
\begin{equation}\label{eq:local:subi}
\benefitq(\Qb_i)=\sum_{q\in \Qb_i} \benefitq(q) \cdot \frac{\sum_{q'\neq q''\in \Qb_i} \frac{\benefitq(q',q'')}{\benefitq(q')+\benefitq(q'')}}{|{\Qb_i \choose 2}|}.\end{equation}






\begin{figure}[!t]\vspace{-3em}
\begin{algorithm}[H]
\setcounter{algocf}{1}
\caption{Multi-Question Selection\label{algorithm:search:centralized}}
\linesnumbered \SetVline
\KwIn{$Q:$ Question Set $Q$; $b$: Selected Question Number}
\KwOut{$Q_b$: Selected Questions}
$P$ = QuestionGrouping($Q$)\;
\ForEach {$P_i \in P $}
{$W_i,W'_i$ = LocalSelection($P_i$, $b$);}
$Q_b$ = GlobalSelection($W$, $W'$ $b$, $|P|$)\;
return $Q_b$\;
\end{algorithm}

\begin{function}[H]
\caption{QuestionGrouping($Q$)}
\linesnumbered \SetVline
\KwIn{$Q:$ Question Set}
\KwOut{$P:$ A set of groups}
\For{$q\neq q'\in Q$}
{
	\lIf{$\benefitq(q,q')\neq\benefitq(q)+\benefitq(q')$}{
		$Corr(q,q')=Y$;
	}
}
Split $Q$ into groups $P_1,P_2, \cdots, P_{|P|}$ such that $q,q'$ are in the same group iff $Corr(q,q')=Y$\;
\end{function}

\begin{function}[H]
\caption{LocalSelection($P_i$, $b$)}
\linesnumbered \SetVline
\KwIn{$P_i$: Correlated groups;
$b$: Selected number}
\KwOut{$W$: $Max(\benefitq/\mathcal{C})$ matrix; $W'$: Local matrix
}
\For{$j\in[1, \min(b,|P_i|)]$}
{
	\lFor{each $j$-size subset $s_j$ of $P_i$}
		{
			Compute $\frac{\benefitq(s_j)}{\costq(s_j)}$\;
		}
		$W[i][j]$$=$$\max_{j=1}^{|P_i|}\frac{\benefitq(s_j)}{\costq(s_j)}$; $W'[i][j]$$=$$\argmax_{j=1}^{|P_i|}\frac{\benefitq(s_j)}{\costq(s_j)}$\;
}
\end{function}


\begin{function}[H]
\caption{GlobalSelection($W$,$S$, $b$, $|P|$)}
\linesnumbered \SetVline
\KwIn{$W$: $Max(\benefitq/\mathcal{C})$ matrix; $W'$: local matrix; $b$: selected question number; $|P|$: Group number}
\KwOut{Selected Questions $Q_b$}
\For {$j \in [1, b]$}
{
    $F[1][j]=W[1][j]$;
    $F'[1][j]=W'[1][j]$\;
}
\For{$i \in [2, |P|] $}
{
    \For{$j \in [1, b] $}
    {
        $F[i][j]=\max_{k=0}^j(W[i][k]+F[i-1][j-k])$\;
        $F'[i][j]$$=$$W'[i][{\argmax_{k=0}^j{W[i][k]+F[i-1][j-k]}}]$\;
    }
}
return $F'$\;
\end{function}
\vspace{-2em}
\end{figure}

\hi{Question Selection.}  To select a $b$-size question set with the largest benefit/cost ratio, a brute-force method enumerates every possible $b$-size subset \Qb and computes $\benefitq(\Qb)$ based on Equation~\ref{eq:global:sum}. However, this method is rather expensive because it needs to enumerate every possible \Qb.  To tackle this issue, we first group the questions \Qall into $|P|$ groups $P_1, P_2, \cdots, P_{|P|}$, and select the local best $j$-size question subset with the largest benefit/cost ratio from each group $P_i$ for $1\leq j \leq b$ and $1\leq i\leq |P|$. Next, we use the local best question sets to generate the global best question set.  Algorithm~\ref{algorithm:search:centralized} shows the pseudo code. 

\hi{(1) Local Selection.} For each group $P_i$, we enumerate every $j$-size subset of $P_i$, compute the benefit based on Equation~\ref{eq:local:subi}, and select the subset with the maximal benefit/cost ratio. Let $W[i][j]$ denote the largest benefit/cost ratio of all $j$-size subsets in $P_i$ and $W'[i][j]$ denote the corresponding $j$-size subset that has the largest ratio. The time complexity of the local selection is $\mathcal{O}({|P_i|\choose b})$ for group $P_i$.

\hi{(2) Global Selection.}  We use a dynamic programming algorithm to select the question set \Qb with the largest benefit/cost ratio. Let $F[i][j]$ denote the largest benefit/cost ratio where we select $j$ questions from the first $i$ groups, and $F'[i][j]$ denote the selected question set from group $P_i$.  

$F[i][j]$ can be computed based on $F[i-1][j-k]$ for $k\in[0,j]$ as follows. If we select $k$ questions from the $i$-th group, we must select $j-k$ questions from the first $i-1$ groups. As questions in different groups have no correlations, we have 
 \begin{equation}
F[i][j]=\max_{k=0}^j(W[i][k]+F[i-1][j-k]),
\end{equation}
\begin{equation}
F'[i][j]=W'[i][{\argmax_{k=0}^j{(W[i][k]+F[i-1][j-k]}}).
\end{equation}

Then $F[|P|][b]$ is the largest benefit/cost ratio and the corresponding selected questions can be generated based on the matrix $F'$ using a traditional backtracking algorithm. The complexity of the global selection is $\mathcal{O}(|P|\cdot b^2)$. As $|P|\leq 3k$, the complexity depends on $k$ and $b$. In practice, $k$ and $b$ are not large, and thus our global ranking method is also scalable.





\subsection{Discussion on Selecting $b$} \label{sec:b:set}

A small $b$ leads to many iterations and the human will be interrupted many times to answer only a few questions in each iteration. Moreover, a small $b$ will miss the correlation amongst different questions. On the contrary, a large $b$ will decrease the number of iterations and cannot use fine-grained human feedback to select questions. To manage this trade-off, we set $b$ to be the size of the largest question group, i.e., $b=\max_{i=1}^{|P|} |P_i|$. 

\section{Experimental Study}
\label{sec:exp}


We conducted experiments to answer the following questions:  
Do our interleaving techniques improve the accuracy of the golden records (GRs)? 
Are the local ranking and global ranking techniques useful for improving GR accuracy? 
Does considering correlation between the questions (\ie batching) help, and how much?

\begin{figure*}[!t]\vspace{-2em}
 \begin{center}
 \subfigure{
     \includegraphics[width=0.95\textwidth]{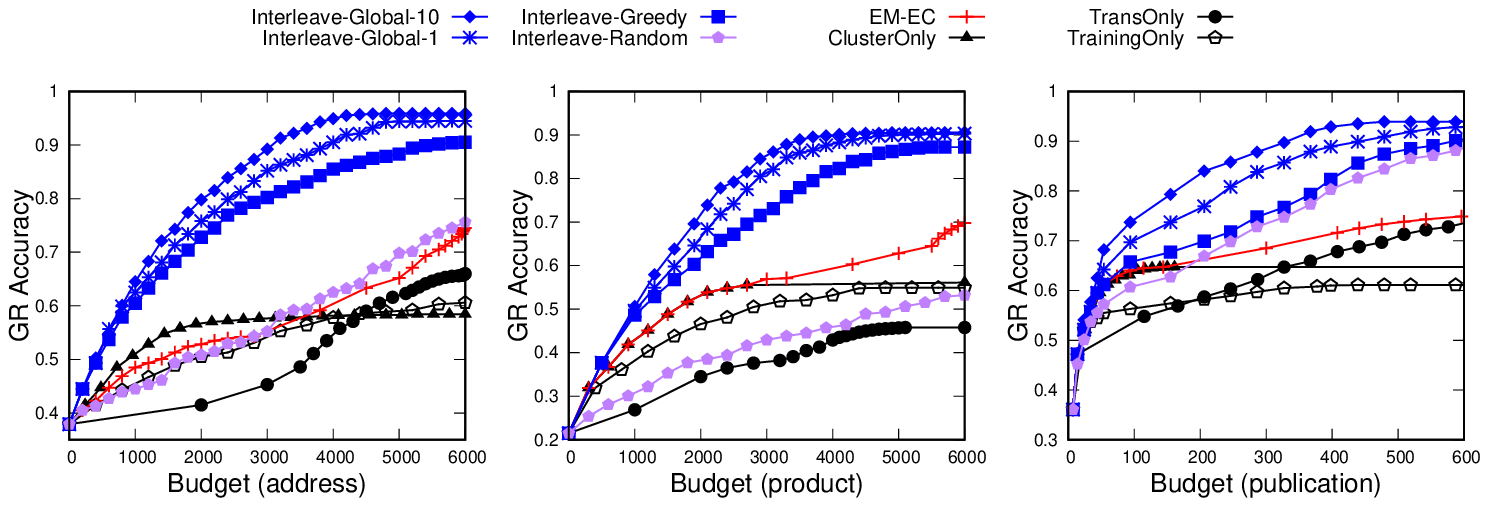}
}
 \end{center}
\vspace{-1.85em}
 \caption{GR Accuracy by Varying Budget (The corresponding maximal numbers of questions are 600, 600, 120)\label{fig:gr:cost}}
\end{figure*}

\vspace{-.5em}
\subsection{Experiment Settings}
\vspace{-.25em}

\begin{table}[!t]\vspace{-2em}
\begin{center}
\caption{Datasets \label{dataset}}
\label{tbl:newdatasets}
\vspace{-1em}
  \begin{tabular}{|l|r|r|r|r|}
    \hline
  Datasets & \product & \addr & \pub \\\hline 
\#Columns  & 6 & 11 & 6 \\\hline
\#Rows & 1,169,376 & 1,040,287 & 120,910 \\\hline
 \#DistinctRows & 191,958 & 140,035 & 11,278 \\\hline
 AvgClusterSize & 6.09 & 7.43 & 10.72 \\\hline
  \end{tabular}
\end{center}
\vspace{-1em}
\end{table}

\hi{Datasets.} We used three real-world datasets (Table~\ref{tbl:newdatasets}). (1) A product dataset \product, where each record is a product. The dataset has 6 columns  (e.g., brand, price, model, type) and 1,169,376 records, and there are 191,958 distinct products.  (2) An address dataset \addr, where each record is the address of a company.  The dataset has 11 columns (e.g., address, city, country, street, latitude, altitude) and 1,040,287 records, and there are 140,035 distinct addresses.  (3) A publication dataset \pub, where each record is a publication.  The dataset has 6 columns (e.g., title, author, journal, volume, year) and 120,910 records, and there are 11,278 distinct publications.  Table~\ref{dataset} shows the statistics of the three datasets.  We manually labeled the ground truth of golden records. We generated the questions as discussed in Section~\ref{subsec:human}. 


\hi{Baselines.} We implemented the following algorithms. All of them first used two blocking rules to generate a set of candidate pairs and then employed different methods to produce the golden records.  
(1) \emec. First run EM, and then if there is little change in the clusters,  switch to EC.  
(2) \qtonly. Only ask training rule questions. 
(3) \qconly. Only ask  cluster questions. 
(4) \qronly.  Only ask  transformation questions. 
(5) \random. Generate local questions and randomly select from them.
(6) \local. First select top-1 questions from each question type, ask these questions to get the answers, and compute the number of changes in golden records for each top-1 question. Suppose question $q$ has the largest number of changes. Then select the next  top-1 question from the question type that $q$ is from. 
(7) \globalone. Select the top-1 question from each set and use the global benefit to select the question with the largest benefit. 
(8) \globalk. Select the top-$k$ questions from each set and use the global benefit to select the question with the largest benefit. 
(9) \globalkcorr. Select the top-$k$ questions from each set and use the global benefit and correlations to select the $b$ questions with the largest benefit.

For EM, we used a random forest based model~\cite{GokhaleDDNRSZ14}. 
For EC, we used the majority voting based method~\cite{DBLP:journals/corr/abs-1709-10436}. 
Our system was implemented in Python. 


\hi{Metrics.} We compared the accuracy of golden records, the F1 score of  the clusters, and the runtime of the algorithms. The cluster precision is the percentage of the computed correct clusters among all computed clusters, the cluster recall is the percentage of the computed correct clusters among all correct clusters, and F1 is the harmonic mean of precision and recall. The cluster F1 is shown in Appendix~\ref{app:exp:cluster}.

\hi{Computing Platform.} 
All experiments were conducted on a Linux server with Intel(R) Xeon(R) E5-2630 2.20GHz CPU, 128GB Memory. We used 20 cores for parallel computing.


\begin{figure}[!t]\vspace{-1.5em}
 \begin{center}
  \hspace*{-1.25em} \subfigure[GR Accuracy]{
     \label{fig:gr:k1:acc}
  \hspace*{-1em} \includegraphics[width=0.26\textwidth]{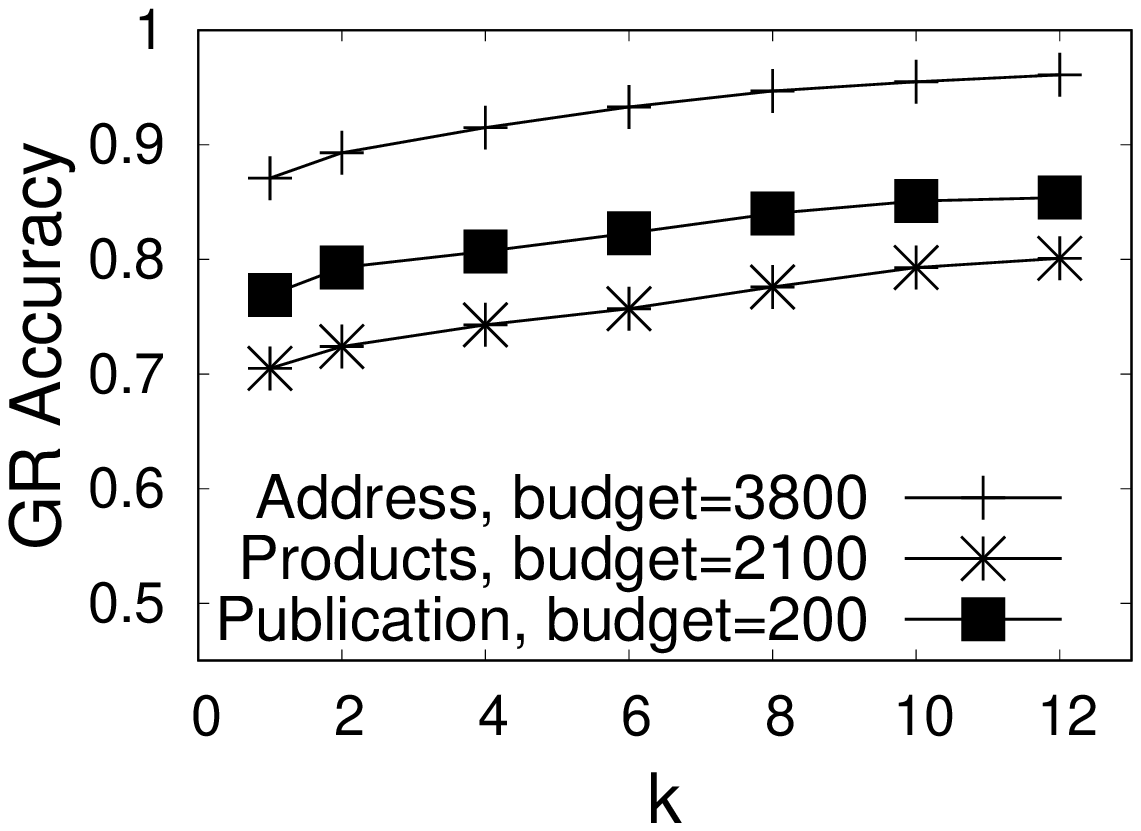}
 }
  \hspace*{-1.25em} \subfigure[Machine Time (\pub)]{
     \label{fig:gr:k1:time}
   \hspace*{-1em}  \includegraphics[width=0.26\textwidth]{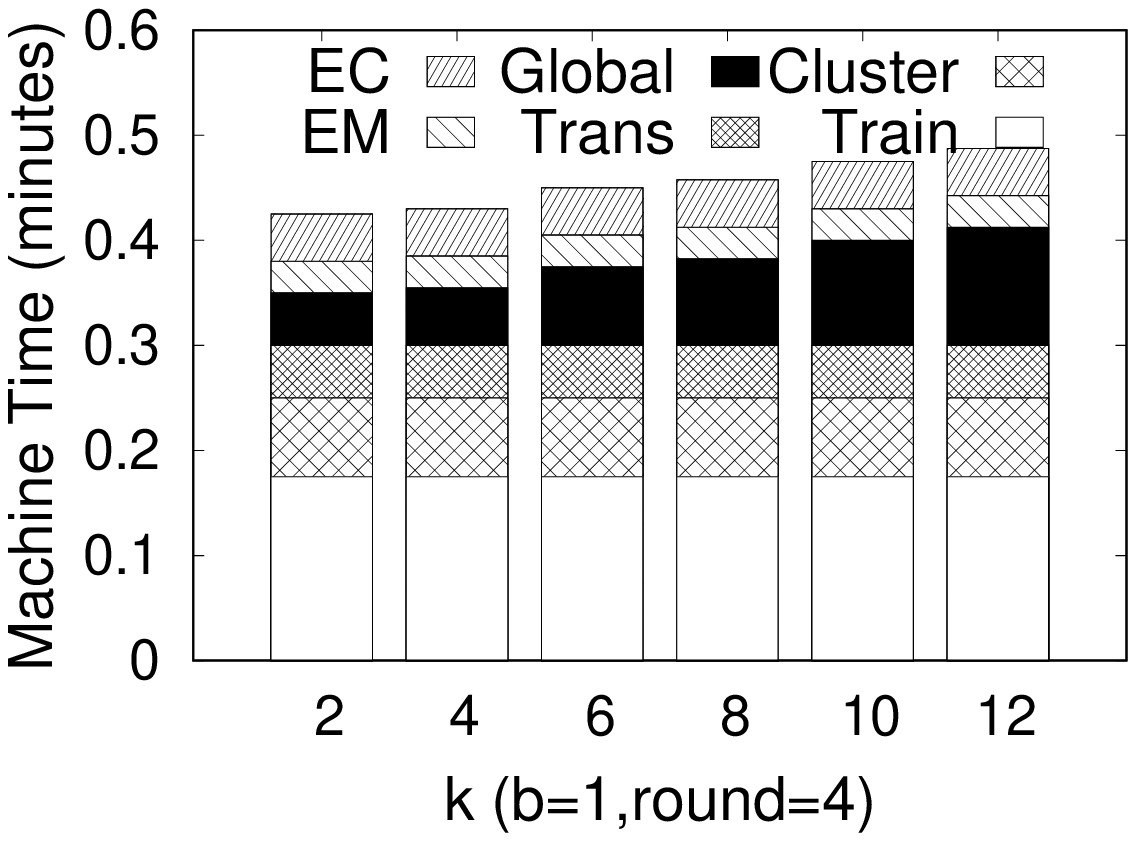}
 }
 \end{center}
\vspace{-1.5em}
 \caption{Varying k\label{fig:gr:k1}}
\vspace{-2em}
\end{figure}

\vspace{-.5em}
\subsection{One Question Per Iteration}
\vspace{-.25em}

We ran the algorithms in multiple rounds. In each round, we selected 10 questions using each algorithm and asked the human to answer these 10 questions. We reported the GR accuracy and machine time for each algorithm.  

\hi{GR accuracy by varying the budget.}  We compared different methods, varying the budget available for asking questions.  Figure~\ref{fig:gr:cost} shows the results. In the figure, the cost corresponds to the human cost computed based on the model in Section~\ref{subsec:cost}. For example, for a value of 1000, we asked about 100 questions. We also show results where we vary the actual number of questions in Appendix~\ref{appendix:gr:qno}. 


We make the following observations from these results:

\begin{figure*}\vspace{-3em}
\begin{center}
\subfigure{
\includegraphics[width=0.95\textwidth]{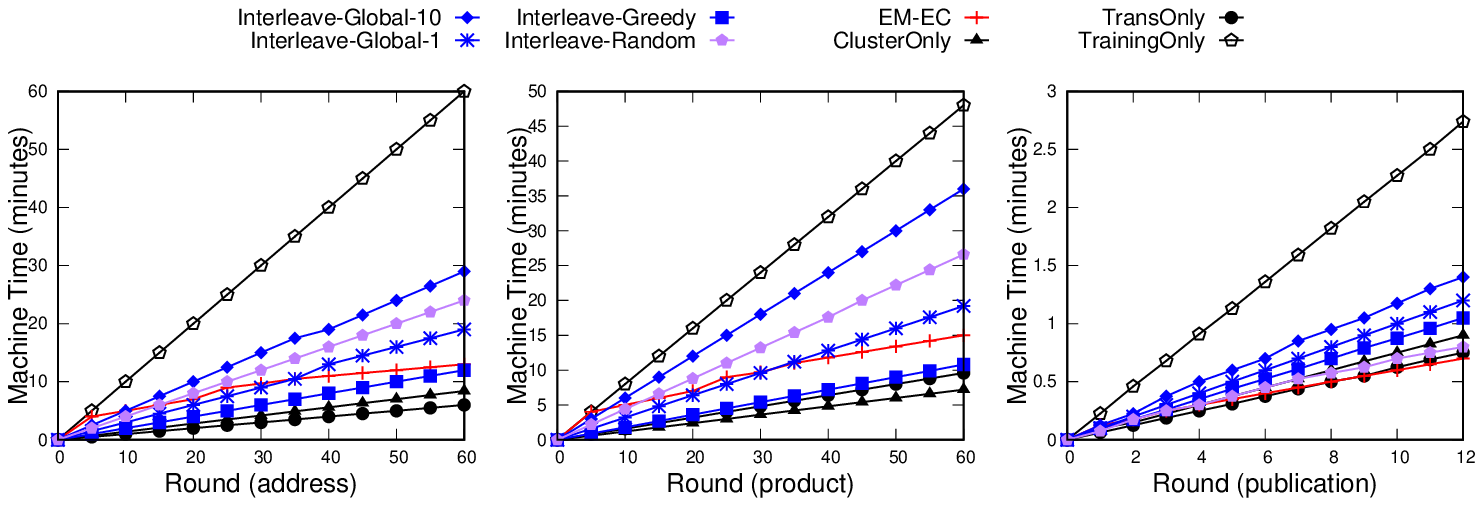}
}
\end{center}
\vspace{-1.5em}
\caption{Runtime by Varying Rounds\label{fig:latency}}
\vspace{0em}
\end{figure*}

{

\noindent (1) The interleaved methods, \globalone, \globalten, \random, and \greedy,  are better than the non-interleaved methods, \emec, \qconly, \qronly, and \qtonly. This is because interleaving questions provides benefits to both EM and EC, demonstrating our key hypothesis that it is essential to allocate human resources holistically across different steps of the data integration pipeline, rather than completing one step at a time.

\noindent (2) Global methods, \globalone and \globalten, that utilize the global benefit to schedule the questions outperform the local methods, \random and \greedy, that only use local rankings to schedule the questions.
This is because the local methods only compare questions within the same type while the global methods use  the global benefit to compare questions from the different types. \greedy is better than \random,  because \random randomly selects questions while \greedy ranks local questions. 
 \globalten outperforms \globalone by 3\%-10\%. This verifies that our global ranking techniques are effective at selecting beneficial questions from different question types. 

\noindent (3)  \greedy outperforms other methods. This verifies that our local ranking techniques are effective at selecting questions of the same type.

\noindent (4)  \qtonly and \qconly achieve lower quality than \emec and \qronly because they require to transform the variant values to the same format.  
\emec is slightly better than \qronly, because besides transforming the variant values, \emec also improves the clustering quality.

\noindent (5)  With the increase of the number of questions, our global method consistently improves the accuracy while other methods do not improve after asking some questions. This is because our methods judiciously select questions, while $(i)$ \emec, \qconly, \qronly and \qtonly only ask a fixed group of questions, and $(ii)$ \random and \greedy only consider local rankings. For example, on the \addr dataset,  our global methods consistently improved up to a cost of 4000 while \qronly, \qconly and \qtonly had stable low quality.

\noindent (6) Although \greedy and \random could reach high accuracy finally, they took  large cost.

\hi{GR accuracy by varying $k$.}  Figure~\ref{fig:gr:k1:acc} shows the results for different $k$. We can see that increasing $k$  improves the accuracy. For example, the GR accuracy is improved by 5\%-8\% when increasing $k$ from 1 to 10. But when $k>10$, the improvement is rather small because the most beneficial global questions are already included in the local top-10 questions. Thus we could set $k=10$.

\hi{Runtime.}  We evaluated the machine time and in each round, we selected 10 questions to ask the human. Figure~\ref{fig:latency} shows the runtime of different methods. The interleaving methods \greedy, \globalone, and \globalten take more time than others, because these three algorithms require us to rank local questions, which is rather expensive.  Note that all methods scale linearly with the  number of answers. \qconly and \qtonly are slower than \qronly, \random and \emec, because (i)~ranking cluster and training questions are more expensive than ranking transformation questions (the former needs to compute the uncertainty while the latter only needs to compute the frequency); and (ii)~\random and \emec do not need to rank questions.  

We computed the time for each component: local ranking, global ranking, EM, and EC. Figure~\ref{fig:gr:k1:time} shows the time for different components on the \addr dataset. We can see that local ranking (training rule ranking, cluster ranking, transformation ranking) takes the bulk of the time, because it needs to compute the uncertainty of many pairs for training questions and cluster questions and identify the frequent transformations. Global ranking takes more time with the larger $k$, as it needs to compare different local questions. Moreover, we use the incremental and parallel computing techniques to improve the efficiency  (see Appendix~\ref{appendix:incremental}).}

\begin{figure*}[!t]\vspace{-2em}
 \begin{center}
 \subfigure[Address]{
     \label{subfig:gr:cor:addr}
 \hspace*{-1em}     \includegraphics[width=0.33\textwidth]{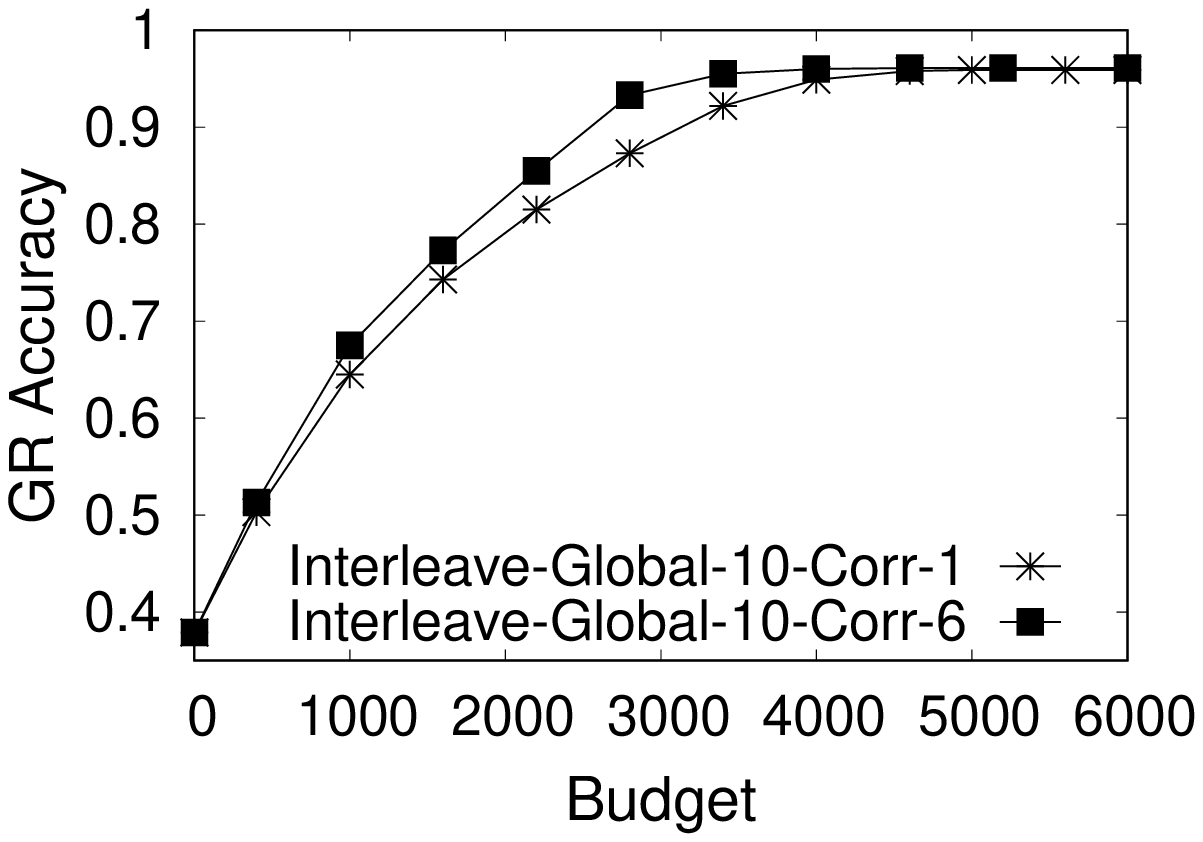}
 }
 \subfigure[Products]{
     \label{subfig:gr:cor:product}
\hspace*{-1em}      \includegraphics[width=0.33\textwidth]{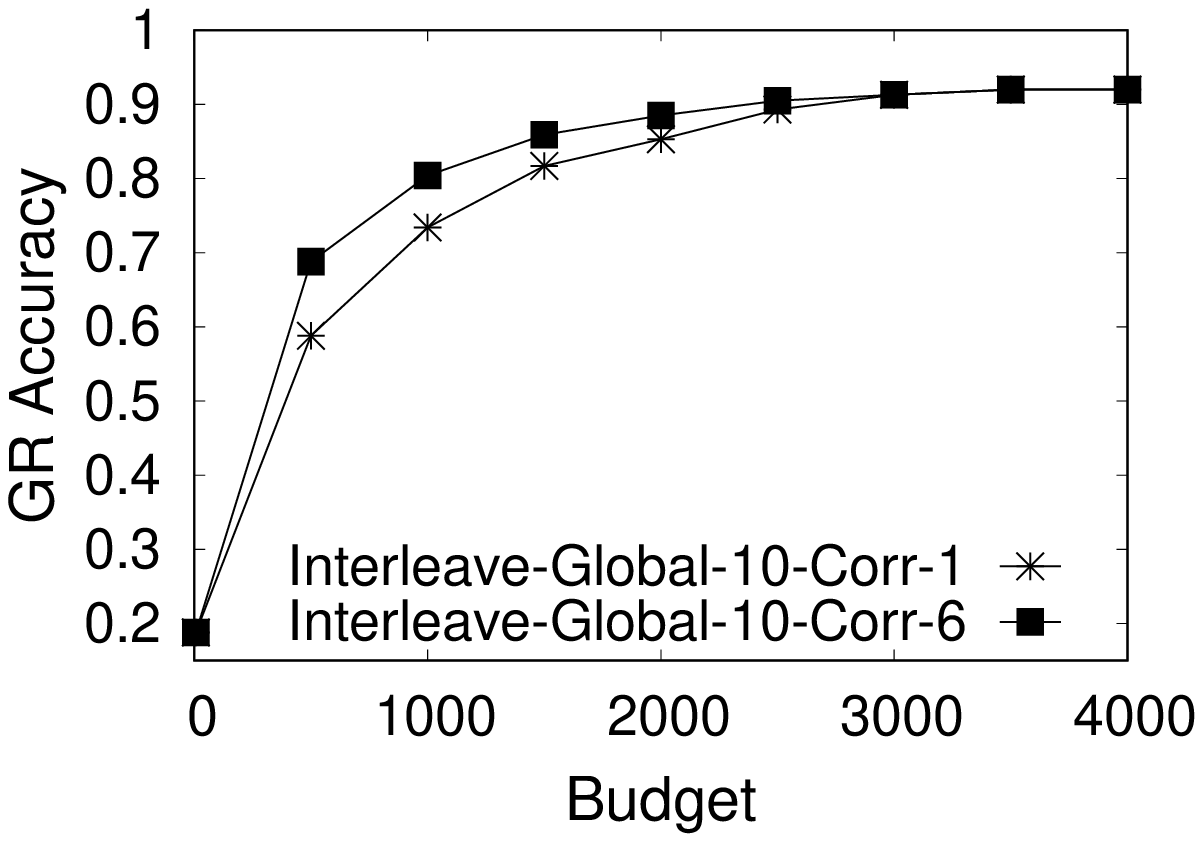}
 }
 \subfigure[Publication]{
     \label{subfig:gr:cor:pub}
\hspace*{-1em}     \includegraphics[width=0.33\textwidth]{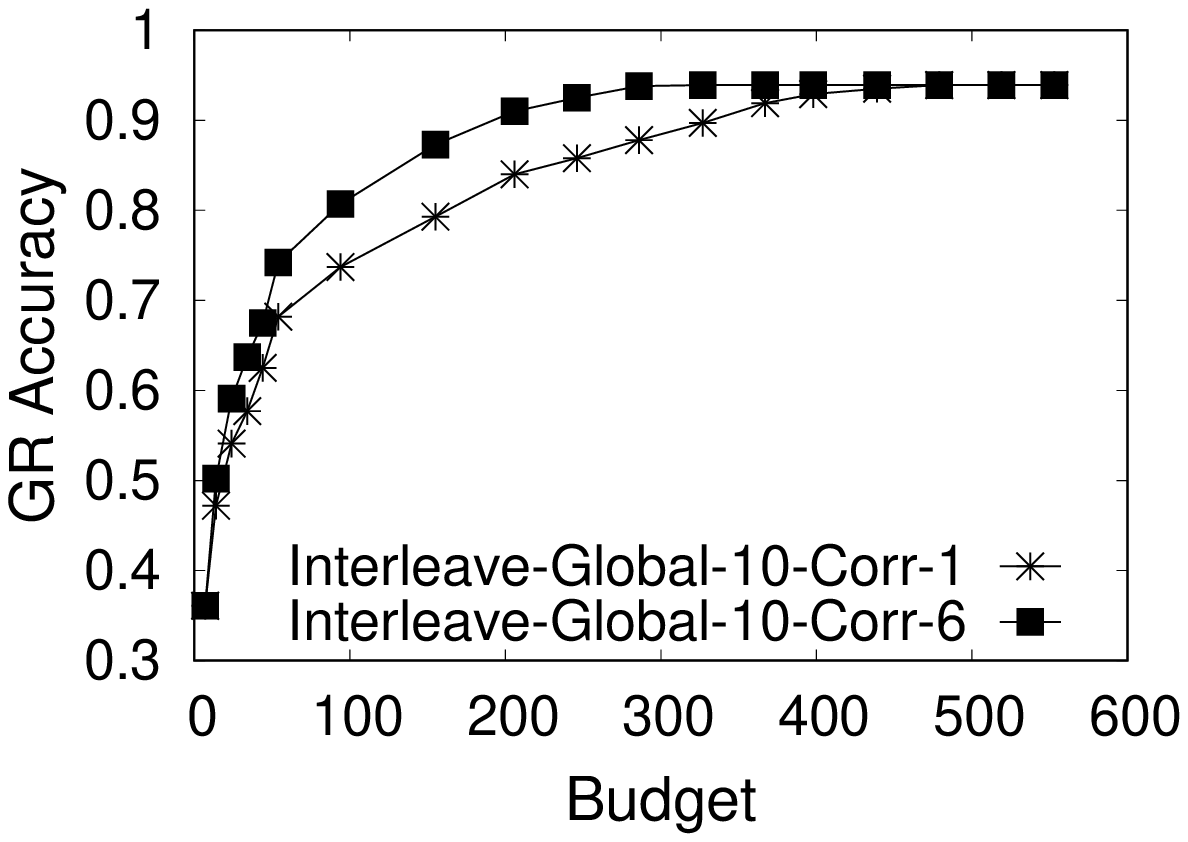}
 }
 \end{center}\vspace{-1.75em}
 \caption{GR Accuracy with Correlation\label{fig:gr:cor}}
\vspace{-1em}
\end{figure*}

\vspace{-.75em}
\subsection{Multiple Questions per Iteration}
\vspace{-.25em}

\begin{figure}[!t]
\vspace{-1.25em}
 \begin{center}
 \subfigure[GR Accuracy]{
     \label{fig:gr:b:acc}
    \hspace*{-1.95em} \includegraphics[width=0.255\textwidth]{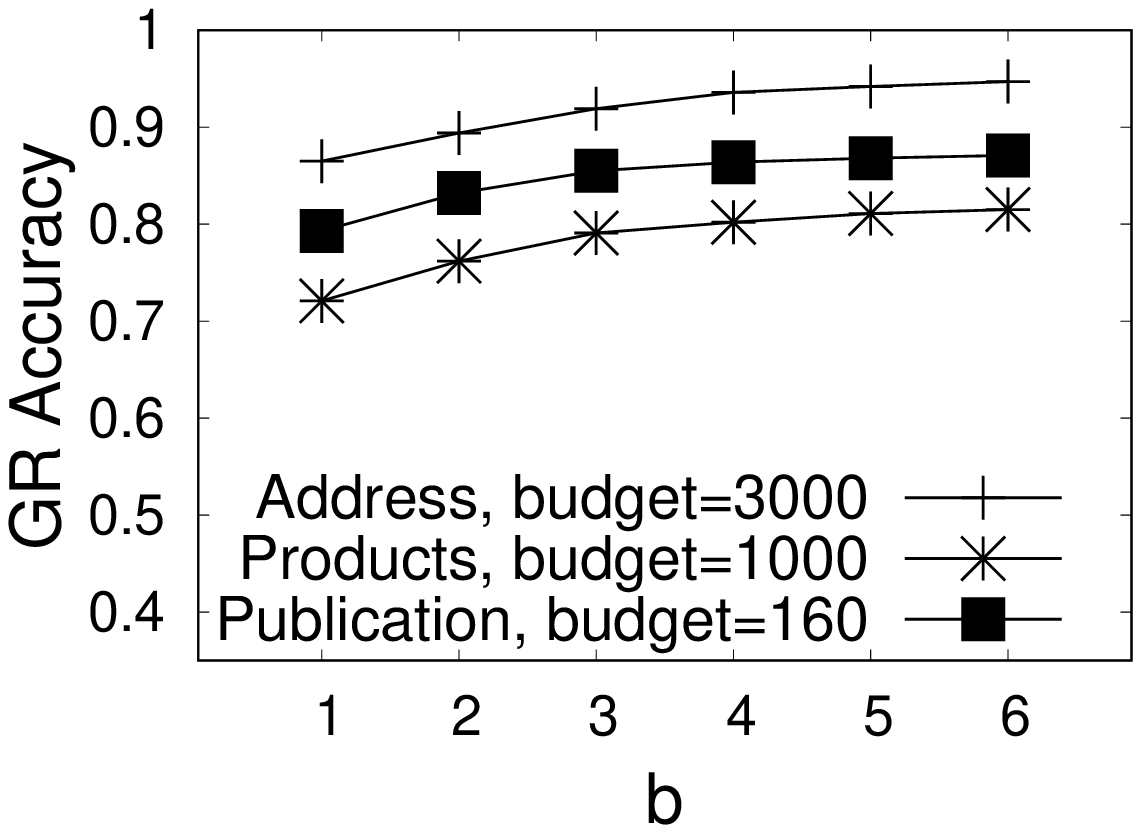}
 }
 \subfigure[Machine Time (\pub)]{
     \label{fig:gr:b:time}
       \hspace*{-1.75em} \includegraphics[width=0.26\textwidth]{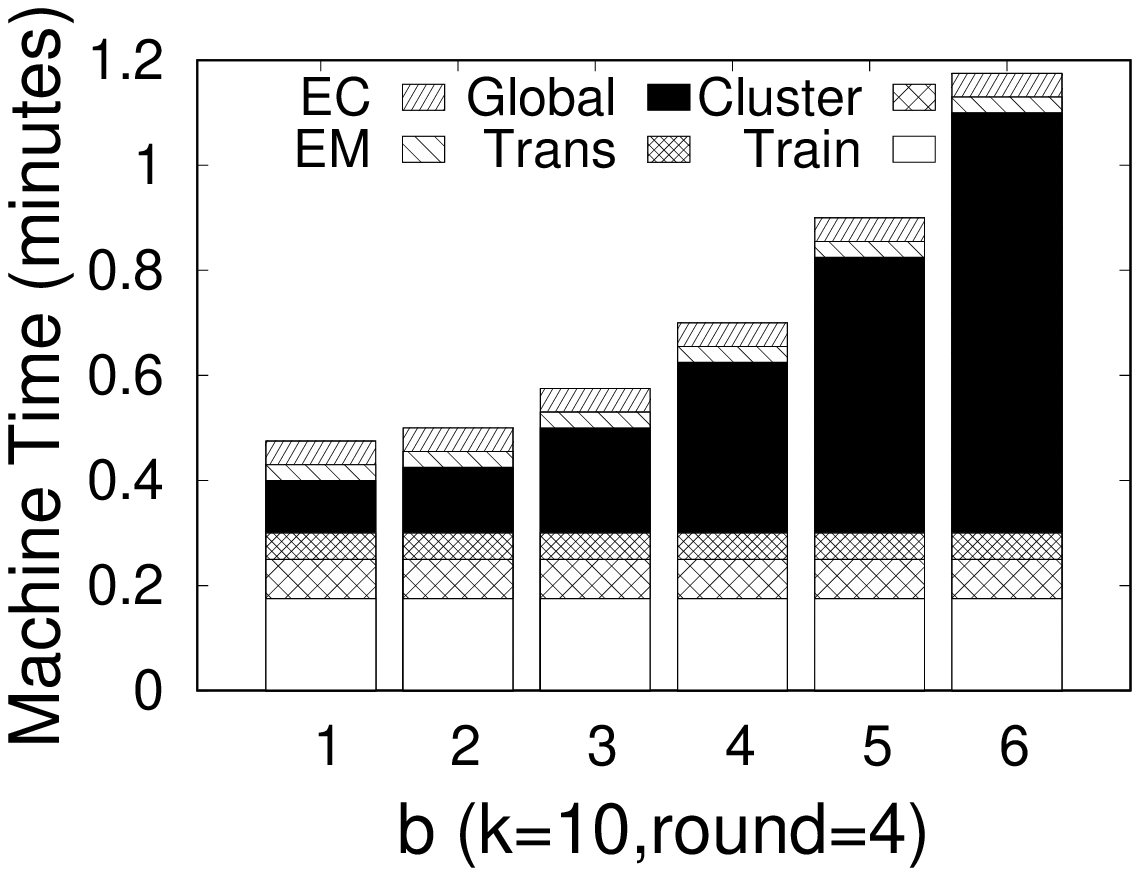}
 }
 \end{center}
 \vspace{-1em}
 \caption{Varying b\label{fig:gr:b}}
\vspace{-2em}
\end{figure}

{\hi{GR accuracy by varying budget.} We compared the batching-based method, \eg the batching based method (selecting $b$ questions per round) to the non-batching method (selecting one question per  round). 
Figure~\ref{fig:gr:cor} shows the results. 
The correlation-based method improves the GR accuracy by 3\%-8\%, because it can use correlation to select the questions and there exists correlations between training/cluster questions and transformation questions. 

\hi{GR accuracy by varying $b$.}  We also varied $b$; Figure~\ref{fig:gr:b:acc} shows the results. As $b$ increases quality improves, because a larger $b$ captures correlations between more questions. After $b>6$ the improvement is small, as only a few questions have correlations. Thus  $b=6$ appears to be a good choice.
 
\hi{Runtime by varying $b$.} Figure~\ref{fig:gr:b:time} shows the running time by varying $b$. With increasing $b$, the running time for global ranking increases, but remains reasonable. For example, from $b=1$ to $b=6$,  global ranking grows from 12 seconds to 60 seconds while the total time grows from 25 seconds to 70 seconds. This is because local ranking depends on the dataset sizes but global ranking depends on $k$ and $b$. 
}

\vspace{-.75em}
\section{Related Work}
\label{sec:rw}
\vspace{-.5em}


\vspace{-.25em}

\hi{Entity Matching.} We can broadly classify EM techniques into three categories: similarity-rule based~\cite{DBLP:journals/pvldb/WangLYF11,DBLP:conf/edbt/PanahiWDN17}, machine-learning-based~\cite{deeper, DBLP:journals/pvldb/KondaDCDABLPZNP16}, 
and crowdsourcing-based\cite{DBLP:journals/pvldb/WangKFF12,DBLP:conf/sigmod/WangLKFF13,DBLP:conf/sigmod/ChaiLLDF16,DBLP:conf/sigmod/DasCDNKDARP17}. The first uses  rules to compute matching pairs. The second first trains a machine-learning model and then utilizes the model to determine matching pairs. The third utilizes human feedback to compute the matching pairs. 

Our method is orthogonal to EM techniques and we can use any EM method to generate the clusters.

\vspace{-.25em}
\hi{Data Transformation.} There are many studies on data transformation for various data types, e.g., numerical data, categorical data and string data~\cite{DBLP:conf/www/HeCCT16,DBLP:conf/sigmod/WangH17,DBLP:conf/sigmod/MorcosAIOPS15,DBLP:journals/pvldb/Singh16,DBLP:conf/icde/AbedjanMIOPS16,DBLP:conf/sigmod/JinACJ17a,DBLP:journals/pvldb/ZhuHC17}. They focus on how to generate the transformation rules using various techniques, e.g., programming synopsis.

Our method is orthogonal to the transformation rule generation techniques and our method could use any existing techniques to generate the transformation rules. 

\vspace{-.15em}
\hi{Entity Consolidation.} The traditional methods on entity consolidation first generate clusters of records and then resolve the conflicts to generate the canonical or preferred values for each attribute~\cite{DBLP:journals/pvldb/LiuDOS11,DBLP:journals/pvldb/ZhengLLSC17,DBLP:journals/pvldb/DongBS09a,DBLP:conf/sigmod/LiLGZFH14,
DBLP:journals/pvldb/LiLGSZDFH14,DBLP:journals/pvldb/GaoLZFH15,DBLP:journals/pvldb/DongN09}. 

Our method is orthogonal to the entity consolidation techniques.  We interleave entity matching and entity consolidation to improve the GR accuracy. We schedule different types of human feedbacks to compute golden records.

\hi{Holistic Data Integraion.}
AnHai et al.~\cite{DBLP:conf/sigmod/DoanABDGKLMPCZ17} pointed out that, in practice, data integration is often an iterative process that heavily involves human-in-the-loop. This requires a substantial extension for current solutions to have more expressive UIs (or questions), as well as effective communications with humans. Their initial result~\cite{DBLP:conf/edbt/PanahiWDN17} made a first attempt on this direction to allow users to manually refine the rules or data via an eyeballing exercise, which cannot be generalized.

Our work makes a major step in filling the above gaps, by having various questions (\ie more expressive UIs) across all components of an end-to-end data integration pipeline, and a smart question scheduling framework (\ie effective communications) to solicit and generalize user feedbacks.



\vspace{-.5em}
\section{Conclusion}
\label{sec:con}
\vspace{-.25em}

In this paper, we  studied the problem of optimizing human involvement in entity matching and consolidation steps to cluster duplicate records and extract a single ``golden record'' for each cluster.  We observed that human input is needed in several areas of this process, including validating clusters, choosing value transformations, and selecting ``golden record'' values.  Our core observation is that by interleaving these questions, rather than asking one type at a time, we can improve overall integration performance.
We then proposed cost models to measure the human time to answer a question through a user study. We introduced two metrics, namely global benefits and local benefits, to evaluate the quality improvement from asking each question type.
We then proposed a question scheduling method that judiciously selects questions to ask, either sequentially (one at a time), or in batches (where we consider correlations on questions), according to a cost budget. Experimental results on real-world datasets showed that our method outperformed existing solutions that only ask one question type at a time, and our system that used batched correlation improved the accuracy from 70\% to 90\%. 

\setcounter{section}{0}
\setcounter{theorem}{0}

\renewcommand\thesection{\Alph{section}}

\appendix

{
\section{Improving Performance} \label{appendix:incremental}

\subsection{Parallel Computing} \label{app:para}


We explain below how each of the tasks in our framework can be easily parallelized.

\hi{Entity Matching.}  Entity matching first gets some candidate pairs and then checks whether these candidate pairs are matching using the trained model. The candidate pairs can be computed offline, and checking whether they are matching can be easily parallelized. Generating the clusters can also be parallelized by grouping the records into partitions, generating the clusters for each partition in parallel and then merging the clusters. 

\hi{Entity Consolidation}. The golden records for different partitions can be inferred in parallel.

\hi{Local Ranking for Training Rule Questions.}  The scores of different training rules can be computed in parallel. 

\hi{Local Ranking for Cluster Questions.}  The scores of different clusters can be computed in parallel.  

\hi{Local Ranking for Transformation Questions.}  We split the records into different partitions, compute the transformation rule for each partition, and then merge the transformation rule. We can then compute the score of different transformations in parallel. 

\hi{Global Ranking.} We compute the global score of each local question generated from the local ranking in parallel for $b = 1$. For $b = 2$, we compute the score for each local question pair in parallel. For $b>2$, we first group the questions and compute the score of each group in parallel.  
}




\begin{figure}[!t]
\begin{center}
\label{}
\includegraphics[width=0.35\textwidth]{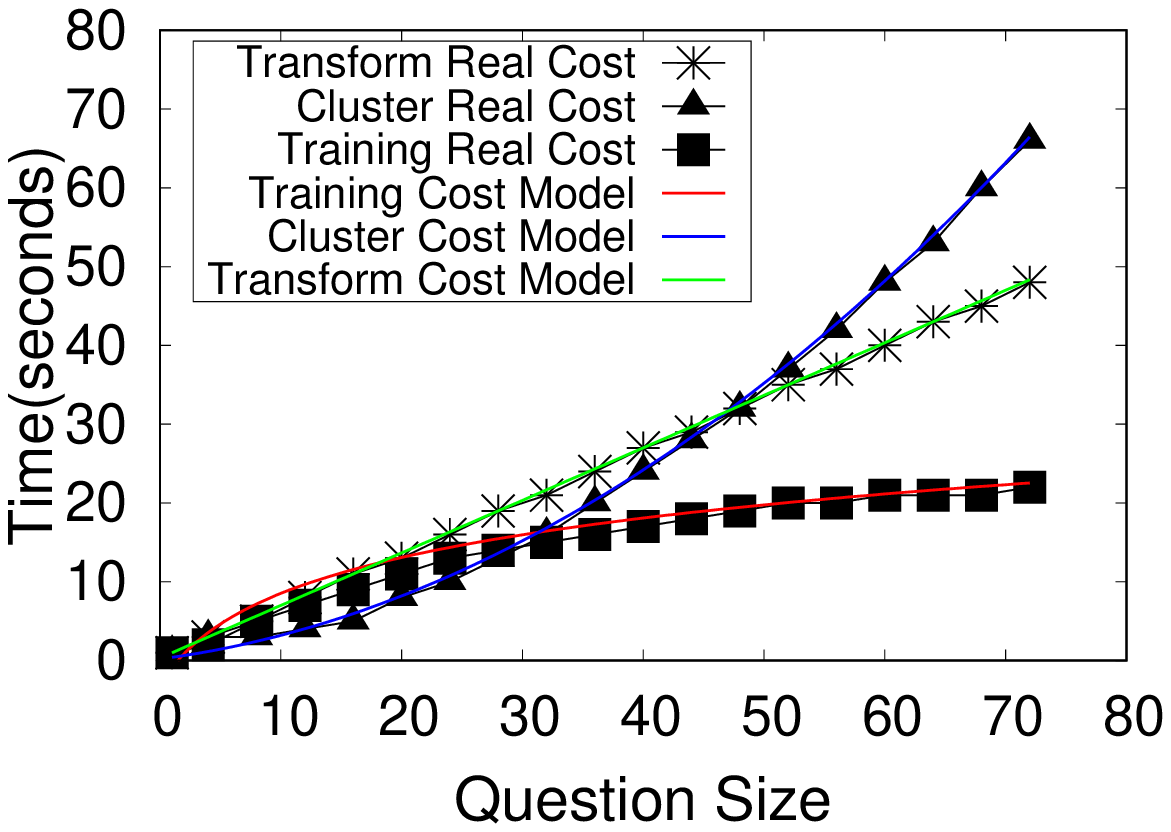}
\vspace{-1em}
\caption{Cost Model via User Study\label{fig:exp:cost:user}}
\end{center}
\end{figure}

\begin{figure*}[!t]
 \vspace{-3em}
 \begin{center}
 \subfigure{
 \label{}
     \includegraphics[width=1.0\textwidth]{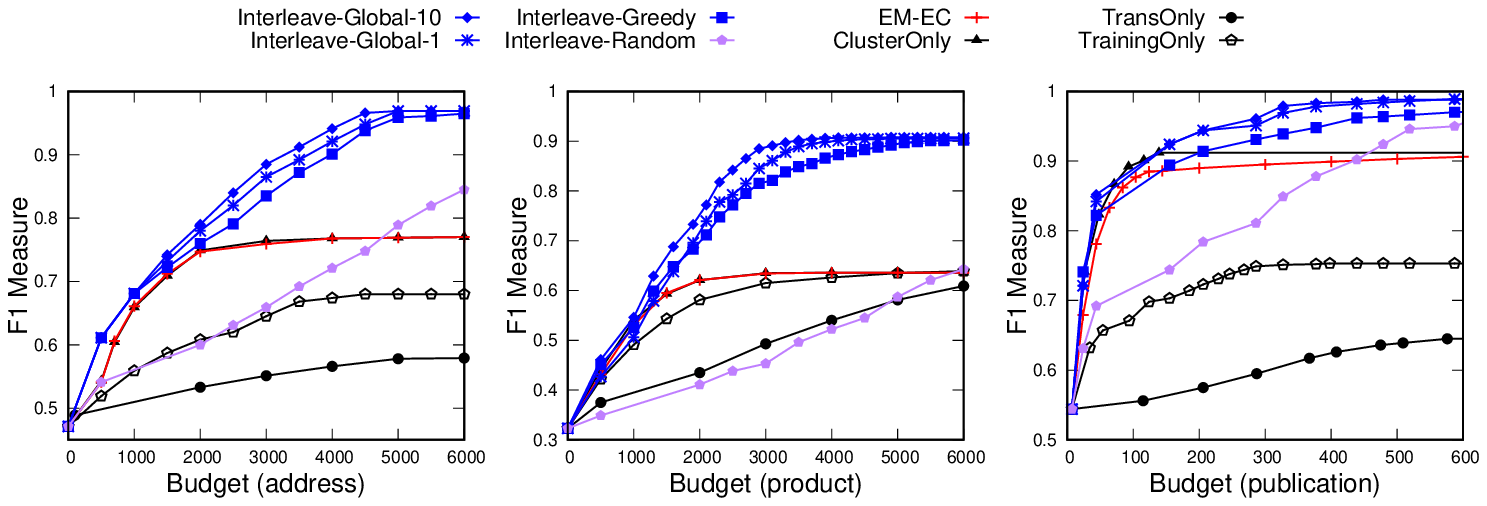}
}
 \end{center}
  \vspace{-1.5em}
 \caption{Clustering Quality by Varying Budget (F1 Measure of Clusters)\label{fig:f1:cost}}
  \vspace{0em}
 \end{figure*}



\begin{figure}[!t]\vspace{-2em}
 \begin{center}
 \subfigure[Machine Time (Products)]{
     \label{}
 \hspace*{-.5em}  \includegraphics[width=0.23\textwidth]{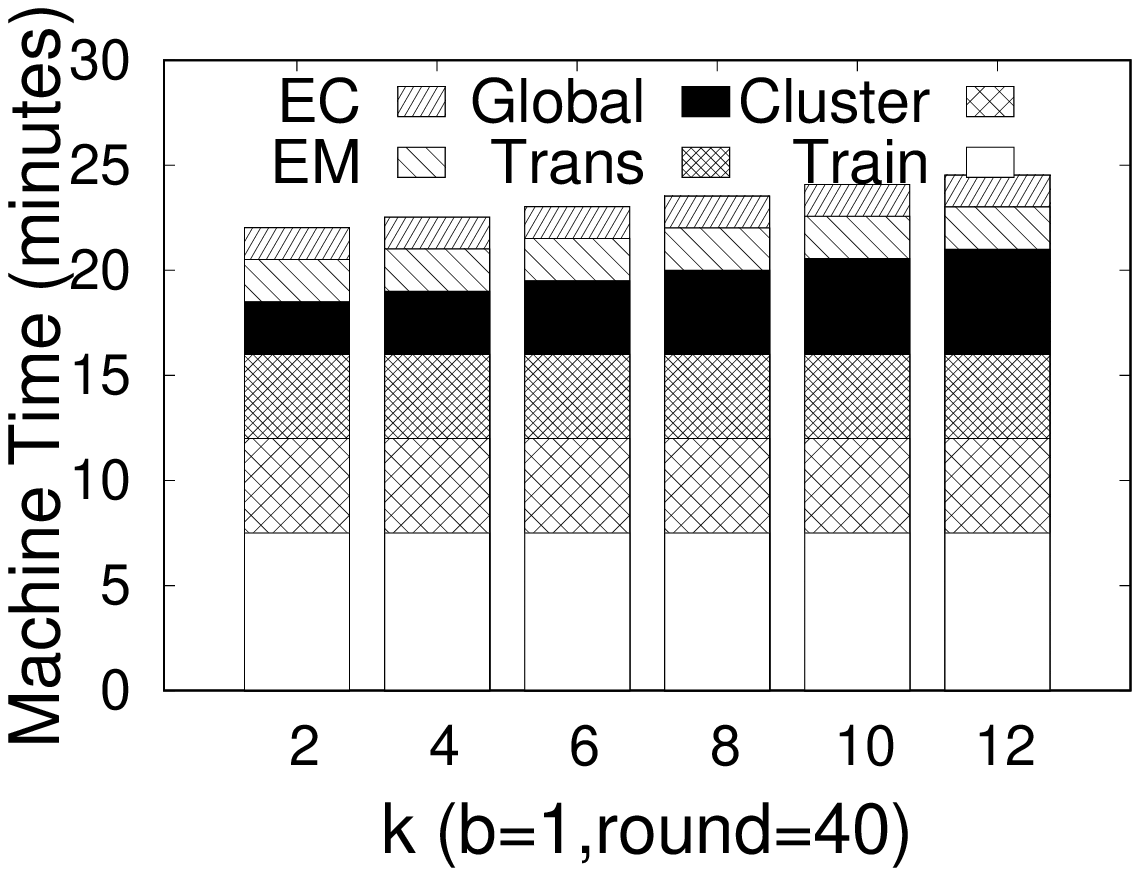}
 }
 \subfigure[Machine Time (Address)]{
     \label{}
    \hspace*{-1em}   \includegraphics[width=0.23\textwidth]{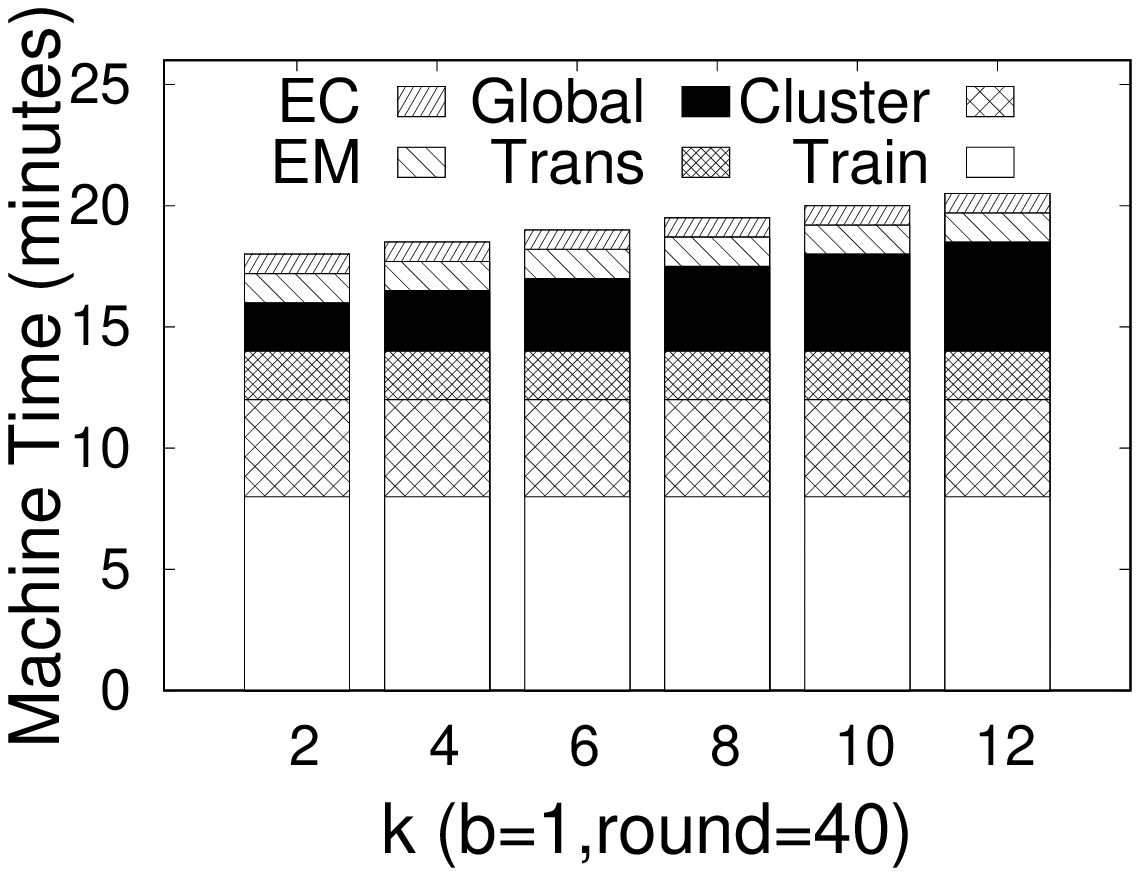}
 }
 \end{center}
  \vspace{-1.5em}
 \caption{Machine Time by Varying k\label{exp:time:k:2}}
 \vspace{-2em}
 \end{figure}

\begin{figure}[!t]
 \begin{center}
 \subfigure[Machine Time (Products)]{
     \label{}
    \hspace*{-.5em}   \includegraphics[width=0.23\textwidth]{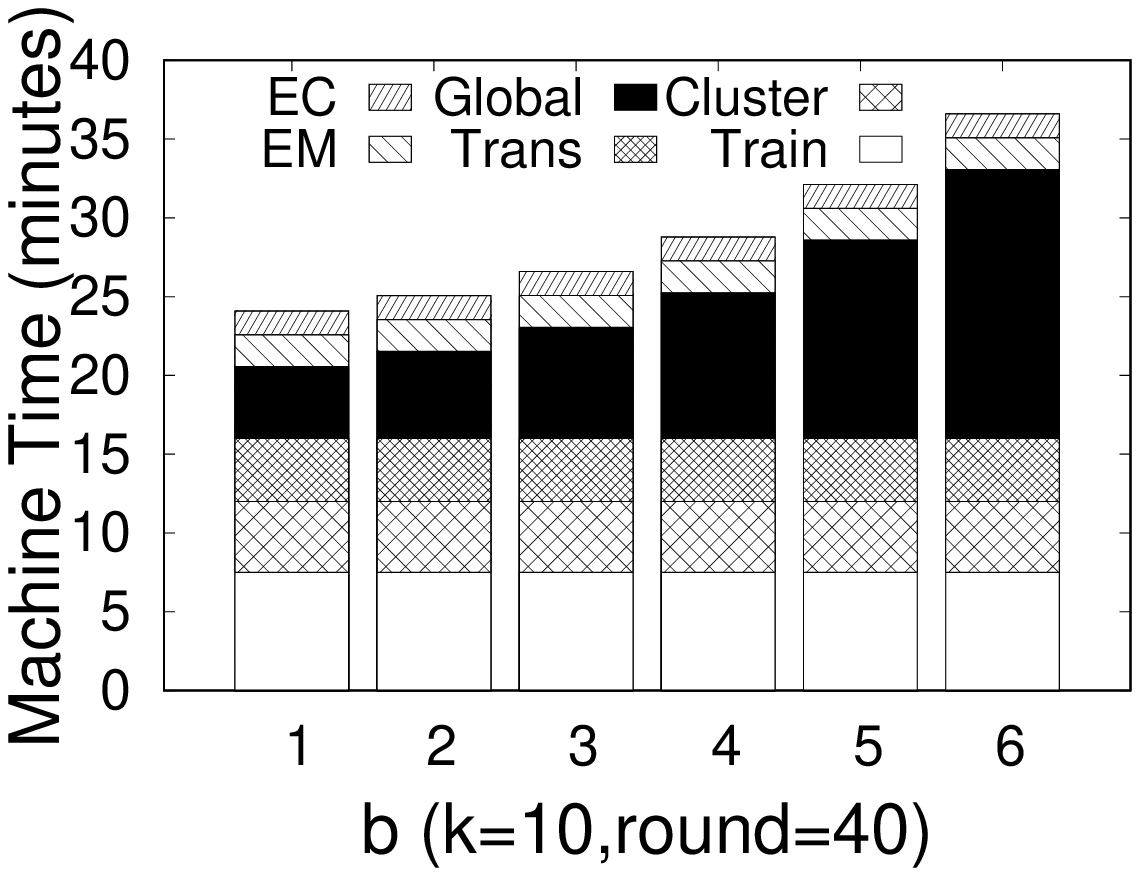}
 }
 \subfigure[Machine Time (Address)]{
     \label{}
 \hspace*{-1em}      \includegraphics[width=0.23\textwidth]{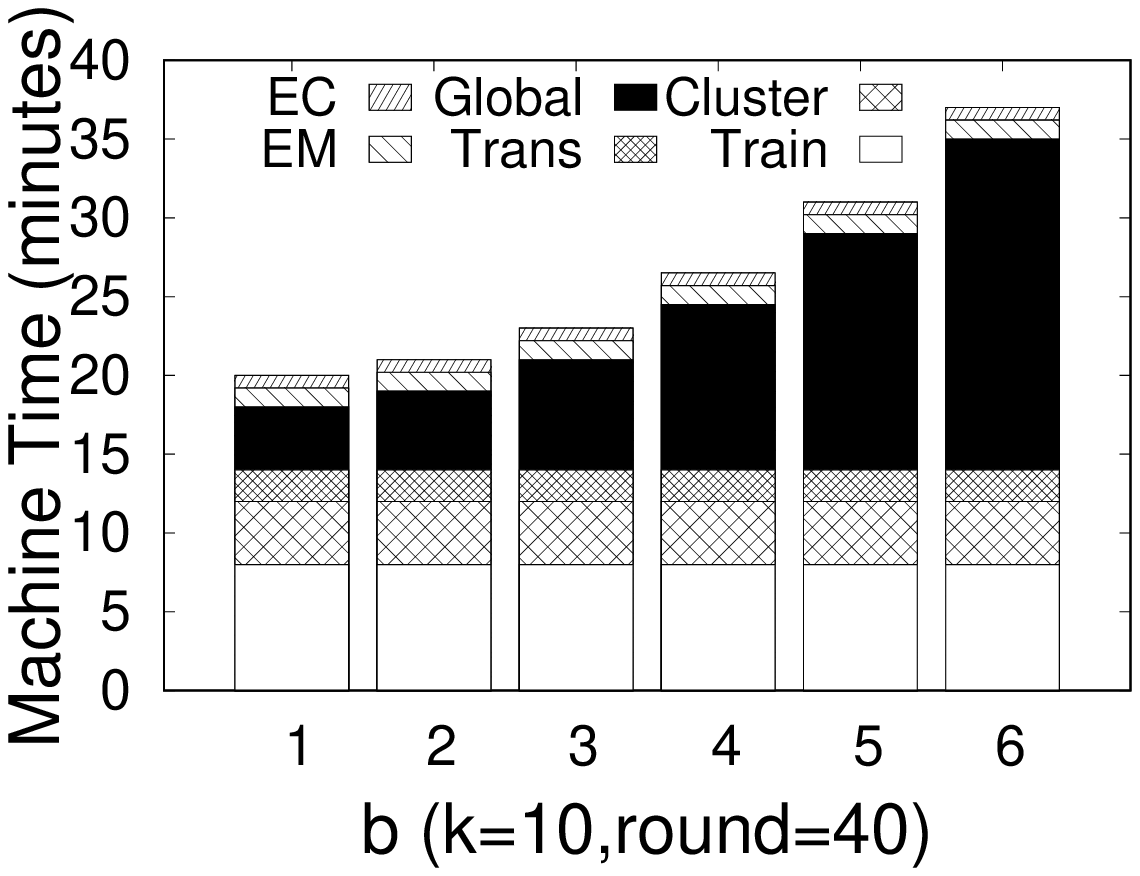}
 }
 \end{center}
  \vspace{-1.5em}
 \caption{Machine Time by Varying b\label{exp:time:b:2}}
  \vspace{-1.5em}
 \end{figure}

{

\subsection{Incremental Computing} \label{app:incre}

We do not need to rerun the machine algorithm from scratch. Instead, we propose incremental computing techniques. 

\hi{Entity Matching.} We do not need to retrain the EM model and recompute the matching pairs using the EM model from scratch. Instead, we propose an incremental method. First, if some new training data are generated by training questions and cluster questions, we utilize the new training data to update the EM model. Second, 
we consider three cases to recompute the matching pairs. (1) If some records are updated and the EM model is not updated, we first incrementally recompute the candidate pairs $(r,r')$, where $r$ is an updated record, and $r'$ is any record. Then we utilize the EM model to identify matching pairs from the candidate pairs. In addition, we also keep the original matching pairs that do not contain an update record. (2) If the records are not updated and the EM model is updated, we only need to check whether the original candidate pairs are matching using the updated EM model. (3) If the records are updated and the EM model is updated, we incrementally compute the new candidate pairs, and check whether the new candidate pairs and original candidate pairs  are matching using the EM model.

\hi{Entity Consolidation}. We only need to check  the updated clusters to reproduce the golden records.

 For local ranking, we do not need to compute the score for all training/cluster/transformation questions, and instead we only need to compute the top-$k$ questions for the updated questions. To this end, we use a priority queue to keep the $k$ best questions, maintain an upper bound for these questions, and use the upper bound to prune the other questions that cannot be in the top-$k$ list.  

\hi{Local Ranking for Training Rule Questions.}  If the EM model is updated, we need to recompute the score; otherwise we only need to update the score for the updated records. 

\hi{Local Ranking for Cluster Questions.}  We only need to recompute the scores for the updated clusters.  For top-$k$ computation, we can maintain an upper bound for each cluster, which is the number of pairs in the cluster (as the utility of each pair is at most 1). Consider the minimal score of the current top-$k$ clusters is $\tau$. If the upper bound of a cluster is smaller than $\tau$, we do not need to compute the real score of the cluster as it cannot be in the top-$k$ clusters.

\hi{Local Ranking for Transformation Questions.}  We only need to update the frequency of the transformations.

\hi{Global Ranking.}  We can utilize the similar pruning technique to compute the best global questions. 


\vspace{-.75em}

\section{Supporting Large Clusters} \label{app:large}
\vspace{-.25em}

We propose a hierarchical clustering based method to support large clusters. Given a large cluster, we first split it into $x$ clusters (e.g., $x=10$). Then if the size of a sub-cluster is larger than $x$, we recursively split it into $x$ clusters.  After building the hierarchical structure, we select the leaf node as a question to ask. If a leaf cluster is approved, besides taking each pair of records within the cluster as a matching pair as discussed in Section~\ref{subsec:human}, we delete the leaf and update its parent node by keeping one record as a representative record (removing $x-1$ records in this leaf cluster); otherwise, besides splitting this cluster and generating training data, we delete this node. We recursively select the leaf node as questions to ask until the hierarchical structure is empty. 

The cost of a large cluster question can be computed by the product of the number of nodes in the hierarchical structure and the cost for each node (i.e., a small cluster question).
}




\begin{figure*}[!t]
\vspace{-3em}
 \begin{center}
 \subfigure{
 \label{}
     \includegraphics[width=1.0\textwidth]{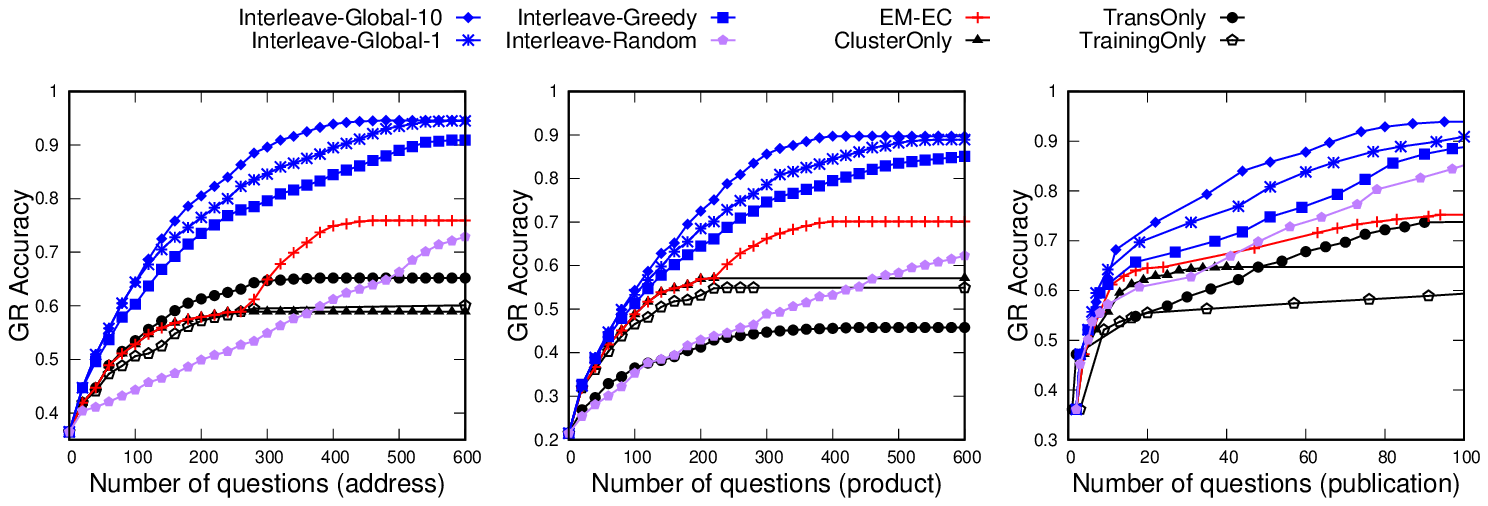}
}
 \end{center}
\vspace{-1.5em}
 \caption{GR Accuracy by Varying Number of Questions  (Accuracy)\label{fig:gr}}
\vspace{-1em}
\end{figure*}

\begin{figure*}[!t]
    \centering
    \begin{minipage}{0.246\textwidth}
        \begin{center}
     \hspace*{-1.5em}    \includegraphics[width=0.9\textwidth]{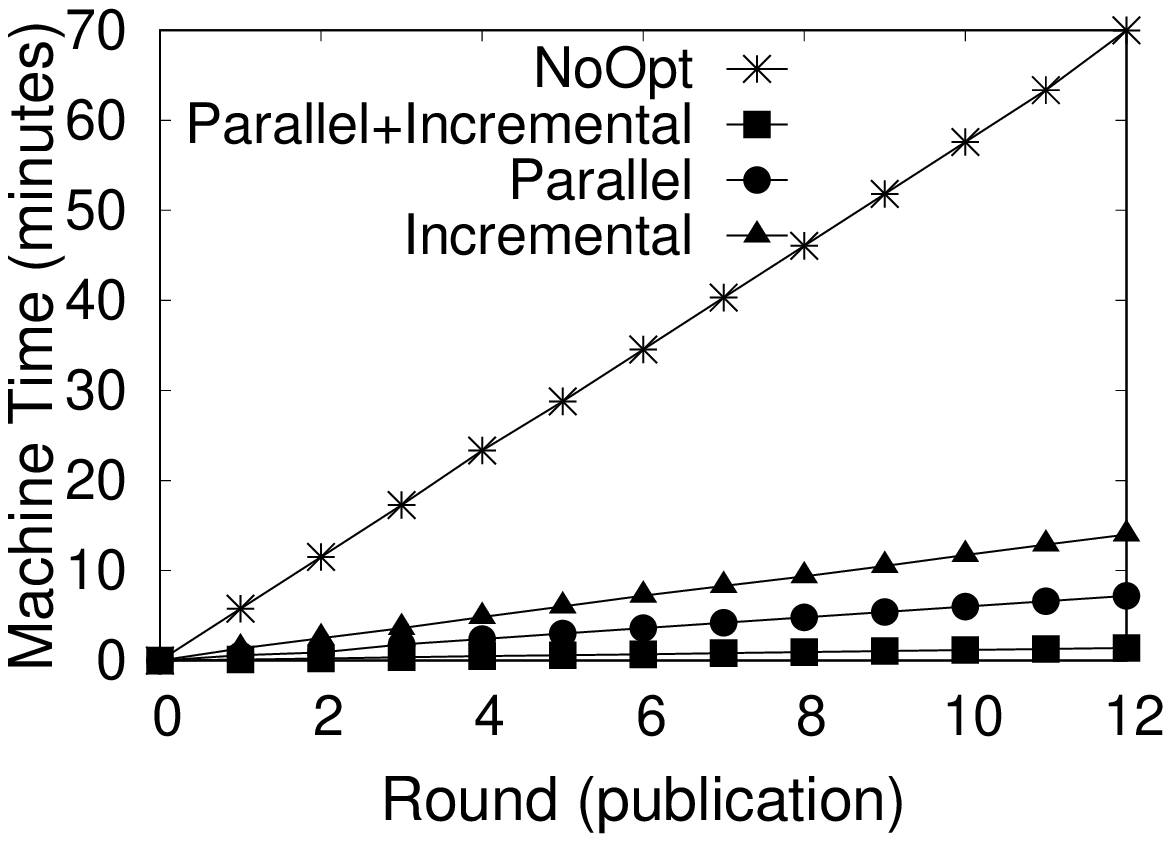}
        \end{center}
        \caption{Incremental and Parallel Computing\label{fig:improve:improve}}
    \end{minipage}%
    \begin{minipage}{0.246\textwidth}
        \begin{center}
       \hspace*{-1.5em} \includegraphics[width=0.95\textwidth]{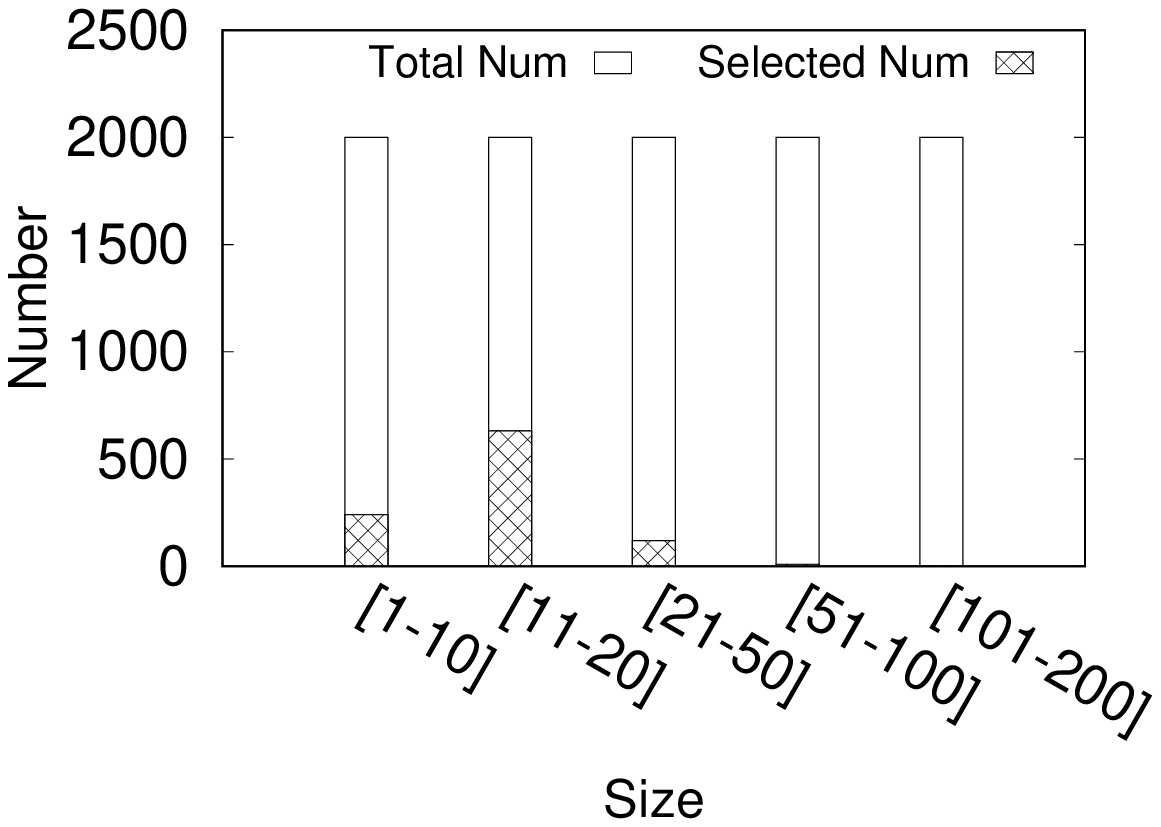}
         \end{center}
         \caption{Number of Selected Clusters\label{fig:gr:size:distribute}}
    \end{minipage}
    \begin{minipage}{0.246\textwidth}
        \begin{center}
      \hspace*{-1.5em}   \includegraphics[width=0.9\textwidth]{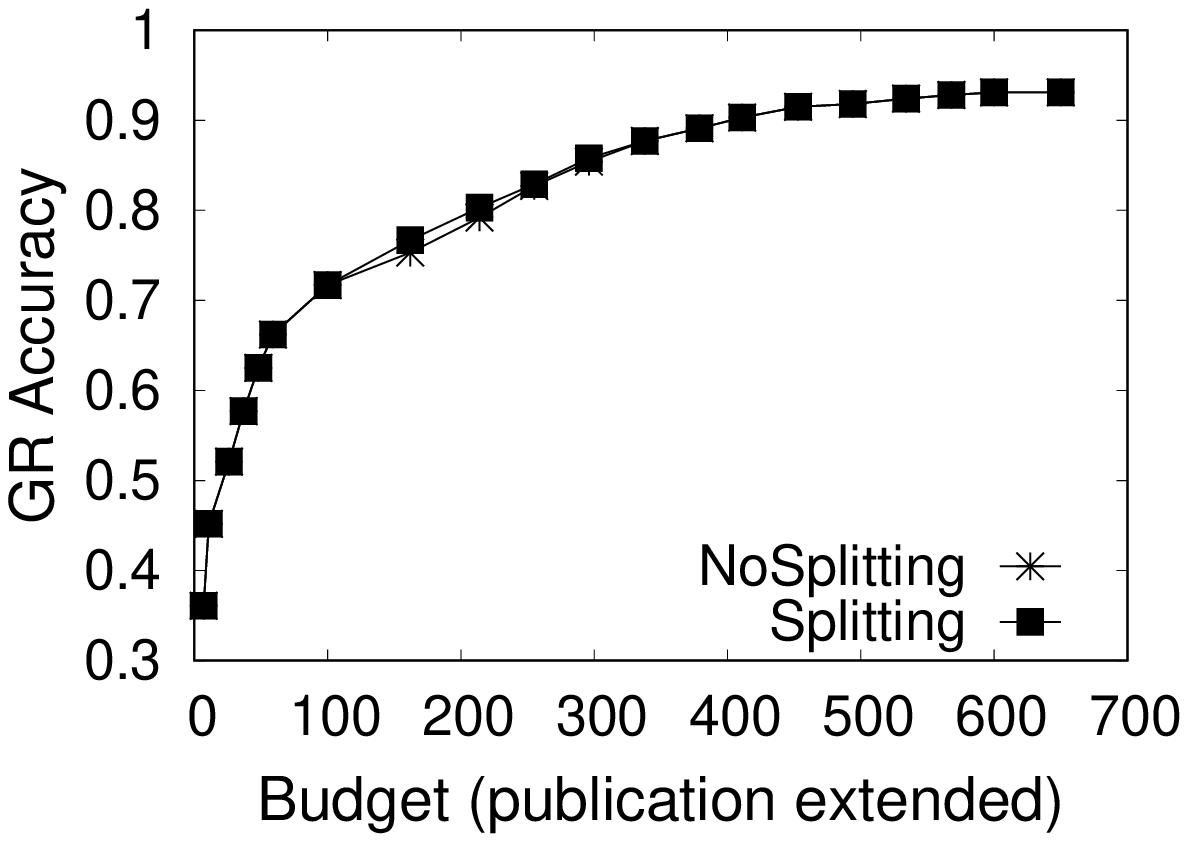}
        \end{center}
        \caption{Splitting Large Cluster\label{fig:gr:cluster}}
    \end{minipage}
    \begin{minipage}{0.246\textwidth}
        \begin{center}
      \hspace*{-1.5em}   \includegraphics[width=0.9\textwidth]{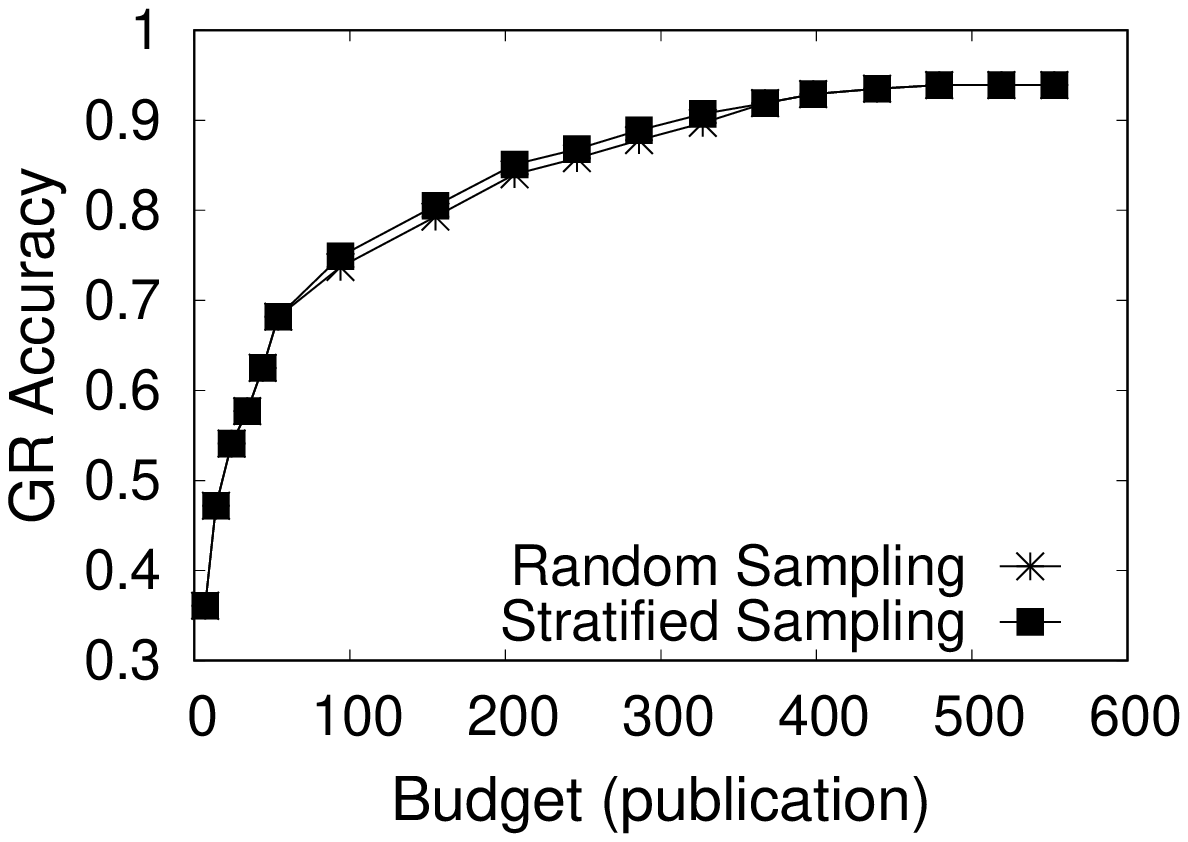}
         \end{center}
         \caption{Sampling Training Question\label{fig:gr:sample}}
    \end{minipage}
\vspace{-1em}
\end{figure*}

\vspace{-.75em}

\section{Complexity Analysis} \label{appendix:complexity}

\vspace{-.25em}

\hi{Complexity of Computing Global Benefit.}  Computing the global benefit of a question includes estimating the probability of its answer and rerunning the EM and EC algorithms to compute the number of changes in the golden records. Estimating the probability for the training rule and transformation is $\mathcal{O}(1)$ while that for the cluster question is $\mathcal{O}(|C|^2)$, where $|C|$ is the size of a cluster $C$. Rerunning the EM and EC algorithms is respectively $\mathcal{O}(|Cand|)$ and $\mathcal{O}(|\mathbb{D}|)$, where $|Cand|$ is the number of candidate matching pairs while $|\mathbb{D}|$ is the dataset size.  Note that in the worst case $|Cand|=|\mathbb{D}|^2$.

\hi{Complexity of Computing Local Benefit.} (1) Training Question \qt. It needs to compute the utility for each record pair that obeys the training rule. The time complexity of computing the utility of a pair is in constant time. Thus the complexity is $\mathcal{O}(|\qt|)$, where $|\qt|$ is the number of pairs that obey the rule. (2) Cluster Question $\qc$. Similar to training questions, it needs to compute the utility of each record pair in the cluster. Thus the complexity is $\mathcal{O}(|\qc|^2)$, where $|\qc|$ is the size of the question. (3) Transformation Question \qr. It only needs to count the frequency of the question. If there is no index, the complexity is $\mathcal{O}(|\mathbb{D}|)$. Usually, when generating the transformations, the frequency is kept associated with the question. Thus the complexity is $\mathcal{O}(1)$. However, it is time consuming to generate the transformation rules~\cite{DBLP:journals/corr/abs-1709-10436}. In practice, we just generate the transformation questions once and use them throughout the algorithm (or recompute them in every several iterations). 

From the complexity analysis, we find that computing local benefits is cheaper than computing global benefits.

\vspace{-.55em}

\section{More Experiments}

\vspace{-.35em}

\subsection{Cost Model via User Study } \label{app:exp:costmodel}

\vspace{-.25em}

We conducted a user study to construct the cost model.  For each question type, we varied the question sizes from 1 to 100. For each question size, we selected 100 questions, assigned each question to 10 students, recorded the time to answer the question, and computed the average time as the human cost for this question size. Next we used regression models to construct the cost model.  We tried different models and Figure~\ref{fig:exp:cost:user} showed the result. We had the following observations. Firstly, the training rule questions followed the logarithmic distribution. Secondly, the cluster questions followed the quadratic polynomial distribution. Thirdly, the transformation questions followed the linear mode, i.e., the cost linearly increased with the increase of the question size. 




\subsection{Evaluation on Cluster Quality}
\label{app:exp:cluster}

We also compared the clustering quality of these algorithms by varying the budget and Figure~\ref{fig:f1:cost} shows the results. We made the following observations. Firstly, our global methods \globalone,  \globalten, \greedy and \random still outperformed other methods, because these four methods judiciously asked a mixed group of questions including both training rule, cluster, and transformation questions, while other methods only asked a fixed group of questions. Secondly, \emec outperformed \qtonly, \qconly and \qronly as \emec asked both training rule and transformation rule questions. Thirdly, \qconly and \qtonly outperformed \qronly, because the former two aimed to generate more training data to improve the clustering quality while  \qronly could only transform the variant values but could not enhance the EM model. \qconly was slightly better than \qtonly, because \qtonly contained incorrect matching pairs while \qconly fixed this issue by asking a human to debug the clusters. Fourthly, \qronly could also improve the clustering quality, because it can transform the data, but the improvement was small because the focus of \qronly was to transform the variant data values. Fifthly, the clustering accuracy of all methods was improved with more questions, because they could utilize human feedback to compute the golden records by either refining the clusters or transforming the variant values.

\vspace{-2em}

\subsection{Machine Time with Varying k and b}

We also measured the effect of varying $k$ and $b$ on the runtime on the \addr and \product datasets. Figure~\ref{exp:time:k:2} shows the results for varying $k$ and Figure~\ref{exp:time:b:2} shows the results for varying $b$. These results are consistent  with those presented earlier on the \pub dataset. Note that as $b$ increases, the fraction of global ranking time increased on these two datasets, because the two data sets are small and local ranking takes less time than global ranking. On the contrary, the \pub dataset is large, and local ranking takes more time than global ranking, because local ranking depends on the dataset size while global ranking depends on $k$ and $b$.

\vspace{-.5em}

\subsection{GR Accuracy by Varying \#Question} \label{appendix:gr:qno}

We have showed GR accuracy by varying budget as different questions had diverse human cost. Here, we assumed that the questions had the same cost and Figure~\ref{fig:gr} showed GR accuracy by varying question numbers. We had the following observations. Firstly, all the methods had the same trend with those varying the budget. Secondly, our global methods \globalone and \globalten still outperformed other methods, because human cost was related to question number. Thirdly, \emec outperformed \greedy within some questions range, because \greedy might overestimate the benefit of some questions. Fourthly, our methods only asked about hundreds questions and achieved 90\% accuracy. For example, on the \addr dataset, \globalten achieved 90\% accuracy when asking 300 questions. 

\vspace{-.75em}
\subsection{Evaluation on Incremental and Parallel Techniques} \label{app:exp:scale}
\vspace{-.25em}

We used 20 threads to improve the efficiency. Figure~\ref{fig:improve:improve} shows the machine runtime.  We can see from the results that the parallel computing and incremental techniques can improve the efficiency. The parallel techniques had a speedup of 10, and the incremental techniques had a speed up of 7. The two techniques had a speedup of 70.

\vspace{-.75em}
\subsection{Evaluation on Large Clusters} \label{appendix:large}
\vspace{-.25em}

Since there were no large clusters in our dataset. We simulated some large clusters based on the error types in the \pub dataset as follows. Given a cluster, we added some records by injecting some errors to the records, e.g., edit errors, transformation errors (e.g., 5, 5th). The largest cluter size was 12 in the dataset. We uniformly generated 6000 clusters with size between 20 and 200. We evaluated the number of selected cluster questions. Figure~\ref{fig:gr:size:distribute} showed the results. We could see that our method did not select large clusters (with size $> 30$), since it was very expensive to validate large clusters.

We also evaluated whether our method worked well for large clusters. We compared two methods: splitting and without splitting large clusters. If we did not split the cluster, we simulated the human to answer the question as we knew the ground truth. Figure~\ref{fig:gr:cluster} showed the results. The two methods achieved similar results as our hierarchical clustering method worked well for large clusters.

\vspace{-.75em}
\subsection{Evaluation on Training Rules} \label{app:rule}
\vspace{-.25em}

We compared two algorithms to select sample pairs for training questions: random sampling and stratified sampling. Figure~\ref{fig:gr:sample} showed the results. We could see that the two methods achieved similar results, because the sampling strategies would not significantly affect the human answer. Moreover, the cluster questions could correct the errors.

\end{document}